\documentclass[12pt]{iopart}
\usepackage{iopams}
\usepackage{graphicx}
\usepackage{graphics}

\begin{document}

\newcommand{\vc}[1]{\mathbf{#1}}

\title{Microscopic calculation of the phonon dynamics of Sr$_{2}$RuO$_{4}$ compared with La$_{2}$CuO$_{4}$}
\author{Thomas Bauer, Claus Falter}
\ead{falter@uni-muenster.de}
\address{Institut f\"ur Festk\"orpertheorie, Westf\"alische
Wilhelms-Universit\"at,\\ Wilhelm-Klemm-Str.~10, 48149 M\"unster,
Germany}
\date{\today}

\begin{abstract}
The phonon dynamics of the low-temperature superconductor
Sr$_{2}$RuO$_{4}$ is calculated quantitatively in linear response
theory and compared with the structurally isomorphic high-temperature
superconductor La$_{2}$CuO$_{4}$. Our calculation corrects for a
typical deficit of LDA-based calculations which always predict a too
large electronic $k_{z}$-dispersion insufficient to describe the
$c$-axis response in the real materials. With a more realistic
computation of the electronic band structure the frequency and
wavevector dependent irreducible polarization part of the density
response function is determined and used for adiabatic and nonadiabatic
phonon calculations. Our analysis for Sr$_{2}$RuO$_{4}$ reveals
important differences from the lattice dynamics of $p$- and $n$-doped
cuprates. Consistent with experimental evidence from inelastic neutron
scattering the anomalous doping related softening of the strongly
coupling high-frequency oxygen bond-stretching modes (OBSM) which is
generic for the cuprate superconductors is largely suppressed or
completely absent, respectively, depending on the actual value of the
on-site Coulomb repulsion of the Ru4d orbitals. Also the presence of a
characteristic $\Lambda_{1}$-mode with a very steep dispersion coupling
strongly with the electrons is missing in Sr$_{2}$RuO$_{4}$. Moreover,
we evaluate the possibility of a phonon-plasmon scenario for
Sr$_{2}$RuO$_{4}$ which has been shown recently to be realistic for
La$_{2}$CuO$_{4}$. In contrast to La$_{2}$CuO$_{4}$ in
Sr$_{2}$RuO$_{4}$ the very low lying plasmons are overdamped along the
$c$-axis.
\end{abstract}


\maketitle

\section{Introduction}\label{SecOne}
The discovery of superconductivity in Sr$_{2}$RuO$_{4}$ by Maeno and
coworkers \cite{Maeno94} has attracted widespread attention partially
because of the structural similarity with the cuprate-based
high-temperature superconductors (HTSC's). While the mother compounds
of the HTSC's are charge transfer insulators, and they usually need to
be doped to become metallic and show superconductivity, in
Sr$_{2}$RuO$_{4}$ low-temperature superconductivity at $T_{c}$ = 1.5 K
condenses from a metallic state that is a strongly two-dimensional
Fermi liquid below about 20 K. The Fermi surface consists of three
weakly corrugated cylindrical sheets $\alpha$ being hole-like and
$\beta$ and $\gamma$ which are electon-like \cite{Bergemann03}. This is
in contrast to La$_{2}$CuO$_{4}$ where only one Fermi sheet exists. The
mechanism of conduction in Sr$_{2}$RuO$_{4}$ at higher temperatures is
an interesting issue because the $c$-axis resistivity shows a broad
maximum at around 130 K and for increasing temperatures the resistivity
starts to decrease. In \cite{Schofield05} such a metallic to
nonmetallic crossover in the $c$-axis resistivity of a highly
anisotropic metal like Sr$_{2}$RuO$_{4}$ with no corresponding feature
in the $ab$-plane properties has been related to the strong coupling
between the electrons and a bosonic mode propagating and polarized in
$c$-direction, like the axial oxygen breathing mode ${\rm O}^{Z}_{z}$,
at the $Z$ point of the Brillouin zone (BZ) to be discussed in
\sref{SecThree}.

Superconductivity in Sr$_{2}$RuO$_{4}$ is thought to be of
unconventional character and thus is intensively discussed in the
literature. The current state of the experiments points to triplet
superconductivity in Sr$_{2}$RuO$_{4}$ in contrast to the $d$-wave
singlet pairing in the HTSC's. Most discussion of the superconducting
mechanism for Sr$_{2}$RuO$_{4}$ is focussed on magnetic fluctuations
and Coulomb repulsion, for a review see \cite{Mackenzie03}. However, a
contribution of electron-phonon interaction (EPI) to pairing may not be
ruled out because the superconducting transition exhibits a clear
isotope effect \cite{Mao03}. In case of the HTSC's there is increasing
evidence that EPI is strong and phonons might play an important role
for the electron dynamics, see \cite{Falter05}, \cite{Bauer08} and
references therein.

Thus, a detailed theoretical study of phonon dynamics in
Sr$_{2}$RuO$_{4}$ and a comparison with the situation in the HTSC's is
desirable. The low-temperature superconductor Sr$_{2}$RuO$_{4}$
crystallizes in the K$_{2}$NiF$_{4}$ structure (I4/mmm) isostructural
to the HTSC La$_{2}$CuO$_{4}$. An analysis and comparison of the
lattice dynamics in both materials is the main intention of the present
work. For this purpose we compute the phonon dynamics for
Sr$_{2}$RuO$_{4}$ and oppose it with our earlier calculations for
La$_{2}$CuO$_{4}$.

The most characteristic feature of phonon dynamics in the HTSC's is the
strong softening upon doping of the high-frequency oxygen
bond-stretching modes (OBSM) which seems to be generic for $p$- as well
as $n$-doped cuprates, see \cite{Falter05,Bauer08,Falter06,Pint05}. The
strong frequency renormalization and a related increase of the
linewidths observed for these anomalous modes point to a strong
coupling of the phonons to the charge dynamics. A discussion and
theoretical investigations of these phonon anomalies are presented e.g.
in \cite{Falter05,Bauer08,Falter06,Falter93}. In agreement with
inelastic neutron scattering \cite{Braden07} our calculations
demonstrate that the typical OBSM phonon anomalies of the cuprates are
not present in Sr$_{2}$RuO$_{4}$. Moreover the characteristic
$\Lambda_{1}$ branch related to the strongly coupling ${\rm O}^{Z}_{z}$
mode in La$_{2}$CuO$_{4}$ is missing.

In order to achieve reliable results for the $\Lambda_{1}$ phonon modes
propagating and polarized along the $c$-axis a careful calculation of
the proper polarization part of the density response function is
crucial. This has already been shown for La$_{2}$CuO$_{4}$ in
\cite{Bauer09} where we have proven that the $\Lambda_{1}$ modes are
highly sensitive with respect to the charge response perpendicular to
the CuO plane. The large anisotropy of the electronic structure of the
cuprates is as a rule considerably underestimated in DFT-LDA
calculations and as a consequence the $\Lambda_{1}$ modes are not well
described. So we have modified a LDA-based calculation for
La$_{2}$CuO$_{4}$ to account for the much weaker $k_{z}$-dispersion of
the electronic bandstructure in the real material.

We have optimized the interlayer coupling in such a way that the
significant features of the sensitive $\Lambda_{1}$ phonons are well
described. The same route is followed in the present work for
Sr$_{2}$RuO$_{4}$ to get a realistic representation of the
$\Lambda_{1}$ modes and in particular of ${\rm O}^{Z}_{z}$ which cannot
be described in the ionic shell-model taking additionally homogeneous
electron gas screening into account \cite{Braden07}.

Finally, we investigate the question if there is room for a
phonon-plasmon scenario around the $c$-axis for Sr$_{2}$RuO$_{4}$ which
has been shown to be a realistic option in La$_{2}$CuO$_{4}$
\cite{Bauer09}. Due to the much weaker electronic $k_{z}$-dispersion
obtained in our computations as compared to La$_{2}$CuO$_{4}$, i.e. a
much stronger anisotropy of Sr$_{2}$RuO$_{4}$, the calculated
free-plasmon frequencies along the $c$-axis are about a factor of eight
smaller in the collisionless regime. Damping, generated by interactions
between the electrons as well as interband transitions which are on a
much lower energy scale as for La$_{2}$CuO$_{4}$ very likely leads to
overdamping of the plasmons propagating strictly along the $c$-axis in
Sr$_{2}$RuO$_{4}$. Thus, different from La$_{2}$CuO$_{4}$ coupled
$c$-axis phonon-plasmon modes should not be well defined collective
excitations at least strictly along the $\Lambda \sim (0,0,1)$
direction and accordingly do not enter the list of possible players for
pairing in contrast to La$_{2}$CuO$_{4}$.

The article is organized as follows. In \sref{SecTwo} the theory
necessary to understand the calculated results is shortly reviewed.
\Sref{SecThree} presents the calculations. The modification of a
LDA-based electronic bandstructure providing the basis of the single
particle content of the irreducible polarization part of the density
response function is developed. Furthermore, the phonon dispersion is
calculated in adiabatic approximation for Sr$_{2}$RuO$_{4}$ and a
comparative discussion with the situation in La$_{2}$CuO$_{4}$ is
given. Finally, in a nonadiabatic calculation the possibility of a
phonon-plasmon scenario is examined. A summary of the paper is given in
\sref{SecFour} and the conclusions are drawn.

\section{Theory and modeling}\label{SecTwo}
In the following a brief survey of the theory and modeling is
presented. A detailed description can be found in \cite{Falter93} and
in particular in \cite{Falter99} where the calculation of the coupling
parameters of the theory is presented.

The local part of the electronic charge response and the EPI is
approximated in the spirit of the quasi-ion approach
\cite{Bauer08,Falter88} by an ab initio rigid ion model (RIM) taking
into account covalent ion softening in terms of (static) effective
ionic charges calculated from a tight-binding analysis. The
tight-binding analysis supplies these charges as extracted from the
orbital occupation numbers $Q_{\mu}$ of the $\mu$ (tight-binding)
orbital in question:
\begin{equation}\label{Eq1}
Q_{\mu} = \frac{2}{N} \sum\limits_{n\vc{k}} |C_{\mu n} (\vc{k})|^2.
\end{equation}
$C_{\mu n} (\vc{k})$ stands for the $\mu$-component of the eigenvector
of band $n$ at the wavevector $\vc{k}$ in the first BZ; the summation
in \eref{Eq1} runs over all occupied states and $N$ gives the number of
the elementary cells in the (periodic) crystal.

In addition, scaling of the short-ranged part of certain pair
potentials between the ions is performed to simulate further covalence
effects in the calculation in such a way that the energy-minimized
structure is as close as possible to the experimental one
\cite{Falter95}. Structure optimization and energy minimization is very
important for a reliable calculation of the phonon dynamics through the
dynamical matrix. Taking just the experimental structure data as is
done in many cases in the literature may lead to uncontrolled errors in
the phonon calculations.

The RIM with the corrections just mentioned then serves as an unbiased
reference system for the description of the HTSC's and can be
considered as a first approximation for the insulating state of these
compounds. Starting with such an unprejudiced rigid reference system
non-rigid electronic polarization processes are introduced in form of
more or less localized electronic charge-fluctuations (CF's) at the
outer shells of the ions. Especially in the metallic state of the
HTSC's the latter dominate the {\em nonlocal} contribution of the
electronic density response and the EPI and are particularly important
in the CuO planes and the RuO plane. In addition, \textit{anisotropic}
dipole-fluctuations (DF's) are admitted in our approach
\cite{Falter99,Falter02}, which prove to be specifically of interest
for the ions in the ionic layers mediating the dielectric coupling and
for the polar modes. Thus, the basic variable of our model is the ionic
density which is given in the perturbed state by
\begin{equation}\label{Eq2}
\rho_\alpha(\vc{r},Q_\lambda, \vc{p}_\alpha) =
\rho_\alpha^0(r) + \sum_{\lambda}Q_\lambda \rho_\lambda^{\rm CF}(r)
+ \vc{p}_\alpha \cdot
\hat{\vc{r}} \rho_\alpha^{\rm D}(r).
\end{equation}
$\rho_\alpha^0$ is the density of the unperturbed ion, as used in the
RIM, localized at the sublattice $\alpha$ of the crystal and moving
rigidly with the latter under displacement. The $Q_\lambda$ and
$\rho^{\rm CF}_\lambda$ describe the amplitudes and the form-factors of
the CF's and the last term in equation \eref{Eq2} represents the
dipolar deformation of an ion $\alpha$ with amplitude (dipole moment)
$\vc{p}_\alpha$ and a radial density distribution $\rho_\alpha^{\rm
D}$. $\hat{\vc{r}}$ denotes the unit vector in the direction of
$\vc{r}$. The $\rho^{\rm CF}_\lambda$ are approximated by a spherical
average of the orbital densities of the ionic shells calculated in LDA
taking self-interaction effects (SIC) into account. The dipole density
$\rho_\alpha^{\rm D}$ is obtained from a modified Sternheimer method in
the framework of LDA-SIC \cite{Falter99}. All SIC-calculations are
performed for the average spherical shell in the orbital-averaged form
according to Perdew and Zunger \cite{Perdew81}. For the correlation
part of the energy per electron the parametrization given in
\cite{Perdew81} has been used.

The total energy of the crystal is obtained by assuming that the
density can be approximated by a superposition of overlapping densities
$\rho_\alpha$. The $\rho_\alpha^0$ in equation \eref{Eq2} are also
calculated within LDA-SIC taking environment effects, via a Watson
sphere potential and the calculated static effective charges of the
ions into account. The Watson sphere method is only used for the oxygen
ions and the depth of the Watson sphere potential is set as the
Madelung potential at the corresponding site. Such an approximation
holds well in the HTSC's \cite{Falter95,Krakauer98}. Finally, applying
the pair-potential approximation we get for the total energy:
\begin{equation}\label{Eq3}
E(R,\zeta) = \sum_{\vc{a},\alpha} E_\alpha^\vc{a}(\zeta)
+\frac{1}{2}\sum_{(\vc{a},\alpha)\neq(\vc{b},\beta)}
\Phi_{\alpha\beta}
\left(\vc{R}^\vc{b}_\beta-\vc{R}^\vc{a}_\alpha,\zeta\right).
\end{equation}
The energy $E$ depends on both the configuration of the ions $\{R\}$
and the electronic (charge) degrees of freedom (EDF) $\{\zeta\}$ of the
charge density, i.e. $\{Q_\lambda\}$ and $\{\vc{p}_\alpha\}$ in
equation \eref{Eq2}. $E_\alpha^\vc{a}$ are the energies of the single
ions. $\vc{a}$, $\vc{b}$ denote the elementary cells and $\alpha$,
$\beta$ the corresponding sublattices. The second term in equation
\eref{Eq3} is the interaction energy of the system, expressed in terms
of \textit{anisotropic} pair-interactions $\Phi_{\alpha\beta}$. Both
$E_\alpha^\vc{a}$ and $\Phi_{\alpha\beta}$ in general depend upon
$\zeta$ via $\rho_\alpha$ in equation \eref{Eq2}.

The pair potentials in equation \eref{Eq3} can be seperated into
long-ranged Coulomb contributions and short-ranged terms, for details
see e.g. Ref \cite{Falter99}.

The dynamical matrix in harmonic approximation can be derived as
\begin{eqnarray}\label{Eq4}\nonumber
t_{ij}^{\alpha\beta}(\vc{q}) &=
\left[t_{ij}^{\alpha\beta}(\vc{q})\right]_{\rm RIM}\\ &-
\frac{1}{\sqrt{M_\alpha M_\beta}} \sum_{\kappa,\kappa'}
\left[B^{\kappa\alpha}_i(\vc{q}) \right]^{*} \left[C^{-1}(\vc{q})
\right]_{\kappa\kappa'} B^{\kappa'\beta}_j(\vc{q}).
\end{eqnarray}
The first term on the right hand side denotes the contribution from the
RIM. $M_\alpha$, $M_\beta$ are the masses of the ions and $\vc{q}$ is a
wave vector from the first BZ.

The quantities $\vc{B}(\vc{q})$ and $C(\vc{q})$ in equation \eref{Eq4}
represent the Fourier transforms of the electronic coupling
coefficients and are calculated from the energy in equation \eref{Eq3},
i.e.
\begin{eqnarray}\label{Eq5}
\vc{B}_{\kappa\beta}^{\vc{a}\vc{b}} &= \frac{\partial^2
E(R,\zeta)}{\partial \zeta_\kappa^\vc{a} \partial R_\beta^\vc{b}},
\\\label{Eq6} C_{\kappa\kappa'}^{\vc{a}\vc{b}} &= \frac{\partial^2
E(R,\zeta)}{\partial \zeta_\kappa^\vc{a} \partial
\zeta_{\kappa'}^\vc{b}}.
\end{eqnarray}
$\kappa$ denotes the EDF (CF and DF in the present model, see equation
\eref{Eq2}) in an elementary cell. The $\vc{B}$ coefficients describe
the coupling between the EDF and the displaced ions (bare
electron-phonon coupling), and the coefficients $C$ determine the
interaction between the EDF. The phonon frequencies
$\omega_\sigma(\vc{q})$ and the corresponding eigenvectors
$\vc{e}^\alpha(\vc{q}\sigma)$ of the modes $(\vc{q}\sigma)$ are
obtained from the secular equation for the dynamical matrix in equation
\eref{Eq4}, i.e.
\begin{equation}\label{Eq7}
\sum_{\beta,j} t_{ij}^{\alpha\beta}(\vc{q})e_j^\beta(\vc{q}) =
\omega^2(\vc{q}) e_i^\alpha(\vc{q}).
\end{equation}
The Eqs. \eref{Eq4}-\eref{Eq7} are generally valid and, in particular,
are independent of the specific model for the decomposition of the
perturbed density in equation \eref{Eq2} and the pair approximation in
equation \eref{Eq3} for the energy. The lenghty details of the
calculation of the coupling coefficients $\vc{B}$ and $C$ cannot be
reviewed in this paper. They are presented in \cite{Falter99}. In this
context we remark that the coupling matrix $C_{\kappa\kappa'}(\vc{q})$
of the EDF-EDF interaction, whose inverse appears in equation
\eref{Eq4} for the dynamical matrix, can be written in matrix notation
as
\begin{equation}\label{Eq8}
C = \Pi^{-1} + \widetilde{V}.
\end{equation}
$\Pi^{-1}$ is the inverse of the {\em irreducible (proper) polarization
part} of the density response function (matrix) and contains the
kinetic part to the interaction $C$ while $\widetilde{V}$ embodies the
Hartree and exchange-correlation contribution. $C^{-1}$ needed for the
dynamical matrix and the EPI is closely related to the (linear) density
response function (matrix) and to the inverse dielectric function
(matrix) $\varepsilon^{-1}$, respectively.

Only very few attempts have been made to calculate the phonon
dispersion and the EPI of the HTSC's using the linear response method
in form of density functional perturbation theory (DFPT) within LDA
\cite{Savrasov96,Wang99,Bohnen03,Giustino08}. These calculations
correspond to calculating $\Pi$ and $\widetilde{V}$ in DFT-LDA and for
the \textit{metallic} state only. On the other hand, in our microscopic
modeling DFT-LDA-SIC calculations are performed for the various
densities in equation \eref{Eq2} in order to obtain the coupling
coefficients $\vc{B}$ and $\widetilde{V}$. Including SIC is
particularly important for localized orbitals like Cu3d in the HTSC's.
Our theoretical results for the phonon dispersion
\cite{Falter05,Falter06,Falter02}, which compare well with the
experiments, demonstrate that the approximative calculation of the
coupling coefficients in our approach is sufficient, even for the
localized Cu3d states. Written in matrix notation we get for the
density response matrix the relation
\begin{equation}\label{Eq9}
C^{-1} = \Pi(1+\widetilde{V}\Pi)^{-1} \equiv \Pi \varepsilon^{-1},
\hspace{.7cm} \varepsilon = 1 + \widetilde{V}\Pi.
\end{equation}
The CF-CF submatrix of the matrix $\Pi$ can approximatively be
calculated for the metallic (but not for the undoped and underdoped)
state of the HTSC's from a TBA of a single particle electronic
bandstructure. In this case the electronic polarizability $\Pi$ in
tight-binding representation reads:
\begin{eqnarray}\nonumber
\Pi_{\kappa\kappa'}&(\vc{q},\omega=0) = -\frac{2}{N}\sum_{n,n',\vc{k}}
\frac{f_{n'}(\vc{k}+\vc{q})
-f_{n}(\vc{k})}{E_{n'}(\vc{k}+\vc{q})-E_{n}(\vc{k})}
\times \\\label{Eq10} &\times \left[C_{\kappa n}^{*}(\vc{k})C_{\kappa
n'}(\vc{k}+\vc{q}) \right] \left[C_{\kappa' n}^{*}(\vc{k})C_{\kappa'
n'}(\vc{k}+\vc{q}) \right]^{*}.
\end{eqnarray}
$f$, $E$ and $C$ in equation \eref{Eq10} are the occupation numbers,
the single-particle energies and the expansion coefficientes of the
Bloch-functions in terms of tight-binding functions.

The self-consistent change of an EDF at an ion induced by a phonon mode
$(\vc{q} \sigma)$ with frequency $\omega_\sigma(\vc{q})$ and
eigenvector $\vc{e}^\alpha(\vc{q}\sigma)$ can be derived in the form
\begin{eqnarray}\label{Eq11}
\delta\zeta_\kappa^\vc{a}(\vc{q}\sigma) & = & \left[-\sum_\alpha
\vc{X}^{\kappa\alpha}(\vc{q})\vc{u}_\alpha(\vc{q}\sigma)\right]
e^{i\vc{q}\vc{R}_\kappa^\vc{a}}\nonumber\\
& \equiv & \delta\zeta_\kappa(\vc{q}\sigma)e^{i\vc{q}\vc{R}^\vc{a}},
\end{eqnarray}
with the displacement of the ions
\begin{eqnarray}\label{Eq12}
\vc{u}_\alpha^{\vc{a}}(\vc{q}\sigma) & = &
\left(\frac{\hbar}{2M_\alpha\omega_\sigma(\vc{q})}
\right)^{1/2}\vc{e}^\alpha(\vc{q}\sigma)e^{i\vc{q}\vc{R}^\vc{a}}\nonumber\\
& \equiv & \vc{u}_\alpha(\vc{q}\sigma)e^{i\vc{q}\vc{R}^\vc{a}}.
\end{eqnarray}
The self-consistent response per unit displacement of the EDF in
equation \eref{Eq11} is calculated in linear response theory as:
\begin{equation}\label{Eq13}
\vc{X}(\vc{q}) = \Pi(\vc{q})\varepsilon^{-1}(\vc{q})\vc{B}(\vc{q}) =
C^{-1}(\vc{q})\vc{B}(\vc{q}).
\end{equation}
The generalization for the quantity $\Pi$ in Eqs. \eref{Eq8} and
\eref{Eq9} needed for the kinetic part of the charge response in the
nonadiabatic regime, where dynamical screening effects must be
considered, can be achieved by adding $(\hbar\omega+i \eta)$ to the
differences of the single-particle energies in the denominator of the
expression for $\Pi$ in equation \eref{Eq10}. Other possible
nonadiabatic contributions to $C$ related to dynamical
exchange-correlation effects and the phonons themselves are beyond the
scope of the present approach. Using equation \eref{Eq9} for the
dielectric matrix, $\varepsilon$, and the frequency-dependent version
of the irreducible polarization part, $\Pi$, according to equation
\eref{Eq10}, the free-plasmon dispersion is obtained from the
condition,
\begin{equation} \label{Eq14}
{\rm det} [\varepsilon_{\kappa\kappa'} (\vc{q}, \omega)] = 0 .
\end{equation}
The coupled-mode frequencies of the phonons and the plasmons must be
determined self-consistently from the secular equation \eref{Eq7} for
the dynamical matrix which now contains the frequency $\omega$
implicitly via $\Pi$ in the response function $C^{-1}$. Such a
nonadiabatic approach is necessary for a description of the interlayer
phonons and the charge-response within a small region around the
$c$-axis as performed in \cite{Falter05,Bauer09} and in the present
paper for Sr$_2$RuO$_4$.

\section{Results and discussion} \label{SecThree}
\subsection{Modification of the LDA-based bandstructure for Sr$_{2}$RuO$_{4}$ and $\Lambda_{1}$ phonons}\label{SecThreeOne}

\begin{figure}
\centering
\includegraphics[]{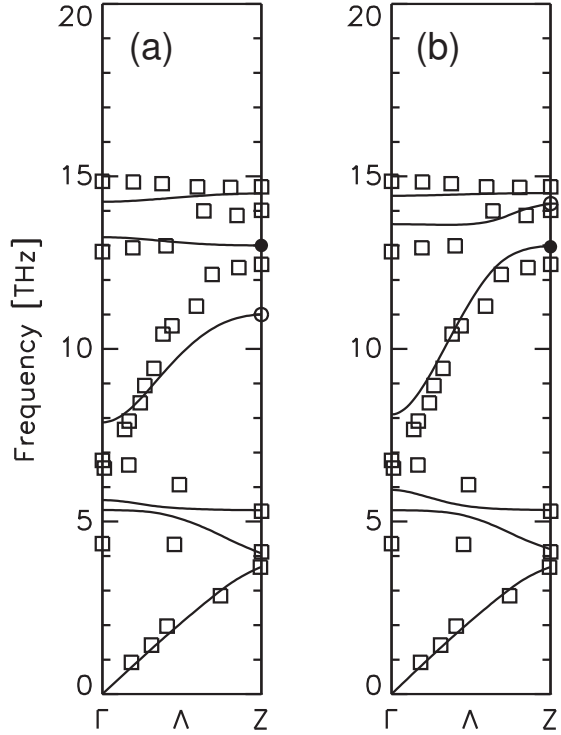}
\caption{Calculated phonon dispersion for La$_{2}$CuO$_{4}$ of the
$c$-axis polarized $\Lambda_1$ modes \cite{Bauer09} based on the 31BM
for the proper polarization part $\Pi_{\kappa\kappa'}$ (a) and the
M31BM (b), respectively. The experimental data are represented as open
squares ($\Box$). The full dot $(\bullet)$ denotes the ${\rm
O}^{Z}_{z}$ mode and the open circle ($\circ$) the $A_{2u}^Z$(ferro)
mode.}\label{Fig_LCO_LAM}
\end{figure}

We shortly recall the construction of a modified LDA-based model for
La$_{2}$CuO$_{4}$ which better describes the real anisotropy of the
HTSC's that as a rule is not correctly represented in LDA being
generally too isotropic. In our effort to obtain a reliable description
of the $\Lambda_{1}$ phonons polarized and propagating along the
$c$-axis of the cuprates \cite{Bauer09} we have shown that these modes
are highly sensitive with respect to details of the $c$-axis coupling
and thus a very accurate representation of the electronic
$k_{z}$-dispersion is needed. Typical LDA based computations lead to an
overestimation of the $k_{z}$-dispersion and give imprecise results for
the $\Lambda_{1}$ modes.

In \fref{Fig_LCO_LAM} we illustrate the degree of inaccuracy for the
$\Lambda_{1}$ phonons, in La$_{2}$CuO$_{4}$ as obtained in
\cite{Bauer09}. \Fref{Fig_LCO_LAM}(a) shows the result of an LDA-based
tight-binding representation of the electronic bandstructure (31 band
model, 31BM) and \fref{Fig_LCO_LAM}(b) gives the outcome of a modified
bandstructure (M31BM) with reduced first neighbour ${\rm O}_{xy}$-La
parameters of the 31BM by 1/6 and of the first neighbour La-La
parameters by 1/3. As seen from \fref{Fig_LCO_LAM} this leads to a much
better result for the phonon modes. The characteristic experimental
features of the dispersion i.e. the step-like structure of the second
highest branch and most significant the third highest branch with the
steep dispersion towards the $Z$ point are not well reflected by the
calculation with the 31BM as input for the proper polarization part in
equation \eref{Eq10}. Using, however, the M31BM the calculated phonon
dispersion is in good agreement with the experiment as can be seen in
\fref{Fig_LCO_LAM}(b). The characteristic features are now well
described.

We also find as a consequence of the increased anisotropy in the M31BM
a rearrangement of the three $Z$ point modes with the highest energy.
While in the 31BM the strongly coupling ${\rm O}^{Z}_{z}$ mode is the
second highest mode and the $A^{Z}_{2u}$(ferro) mode the lowest one in
the M31BM ${\rm O}^{Z}_{z}$ is the lowest and $A_{2u}$(ferro) the
second highest of the three modes. Thus ${\rm O}^{Z}_{z}$ is the end
point of the steep branch. During this mode rearrangement the frequency
of ${\rm O}^{Z}_{z}$ stays nearly constant while that of $A_{2u}^{Z}$
(ferro) is strongly increased.

We have recalled these facts for La$_{2}$CuO$_{4}$ because for
Sr$_{2}$RuO$_{4}$ a comparable high sensibility of the $\Lambda_{1}$
modes in dependence of the electronic $k_{z}$-dispersion occurs and an
analogous mode behaviour for ${\rm O}^{Z}_{z}$ and $A^{Z}_{2u}$(ferro)
results in response to a weaker $k_{z}$-dispersion. Likewise as in
La$_{2}$CuO$_{4}$ this sensitivity of certain $c$-axis phonons is used
to construct an accurate tight-binding description of the electronic
bandstructure (BS) in Sr$_{2}$RuO$_{4}$.

As a first approximation for the BS we employ a tight-binding
representation of the first principles linearized - augmented -
plane-wave (LAPW) BS as obtained within the framework of DFT-LDA
\cite{Mazin00}. This analysis leads to a 27 band model (27BM) including
Sr4d, Ru4d and O2p states.

\begin{figure}
\centering
\includegraphics[]{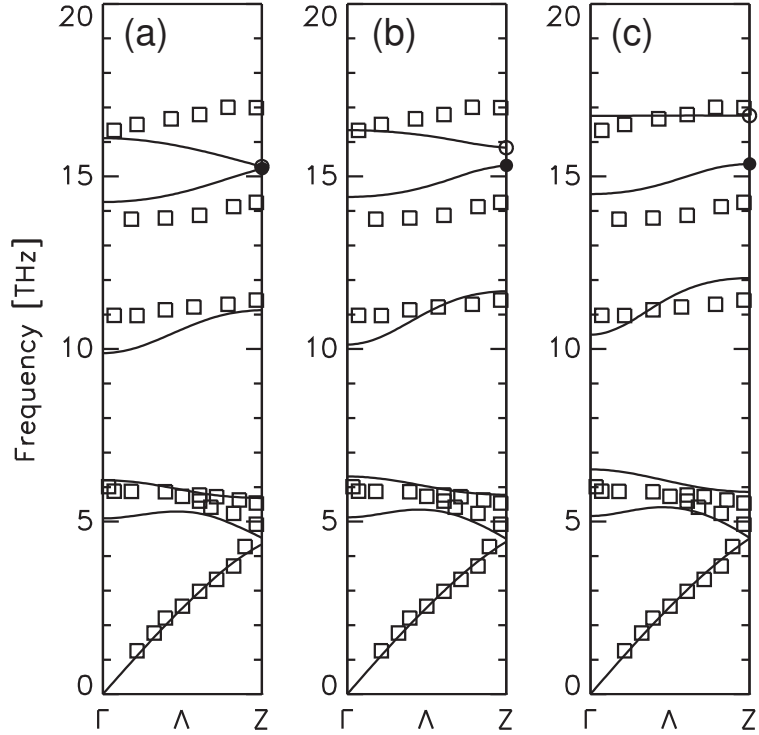}
\caption{Calculated phonon dispersion for Sr$_{2}$RuO$_{4}$ of the
$\Lambda_{1}$ modes based on different models for the electronic
bandstructure as explained in the text. 27BM (a), M27BM1 (b) and M27BM2
(c). The experimental values from \cite{Braden07} are represented as
open squares ($\Box$). The full dot $(\bullet)$ denotes the ${\rm
O}^{Z}_{z}$ mode and the open circle $(\circ)$ the $A_{2u}^Z$(ferro)
mode.}\label{Fig_SRO_LAM}
\end{figure}

\begin{figure}
\centering
\includegraphics[]{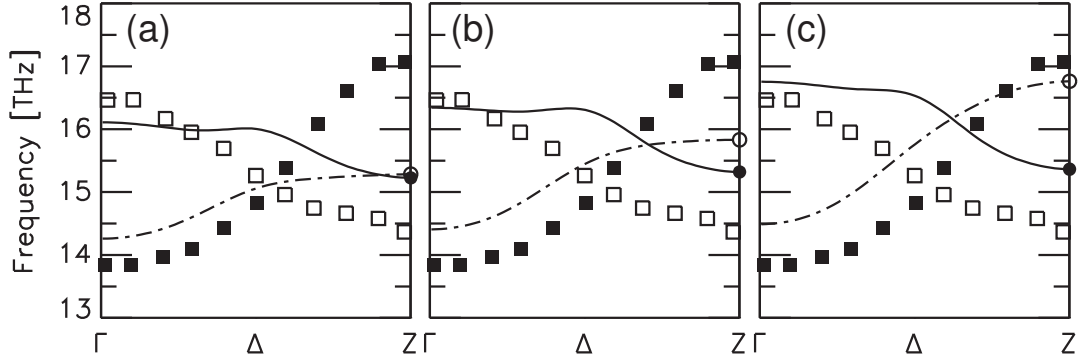}
\caption{Calculated phonon dispersion for Sr$_{2}$RuO$_{4}$ of a
$\Delta_{1}$ mode ($-\!\!-$) and a $\Delta_{4}$ mode ($-\cdot -$) based
on the 27BM (a), the M27BM1 (b) and the M27BM2 (c). The experimental
data points \cite{Braden07} are characterized by open squares ($\Box$)
for $\Delta_{1}$ and by full squares $(\blacksquare)$ for $\Delta_{4}$.
The full dot $(\bullet)$ and the open circle ($\circ$) at the $Z$ point
represent ${\rm O}^{Z}_{z}$ and $A_{2u}^{Z}$,
respectively.}\label{Fig_SRO_DEL1_4}
\end{figure}

\begin{figure}
\centering
\includegraphics[]{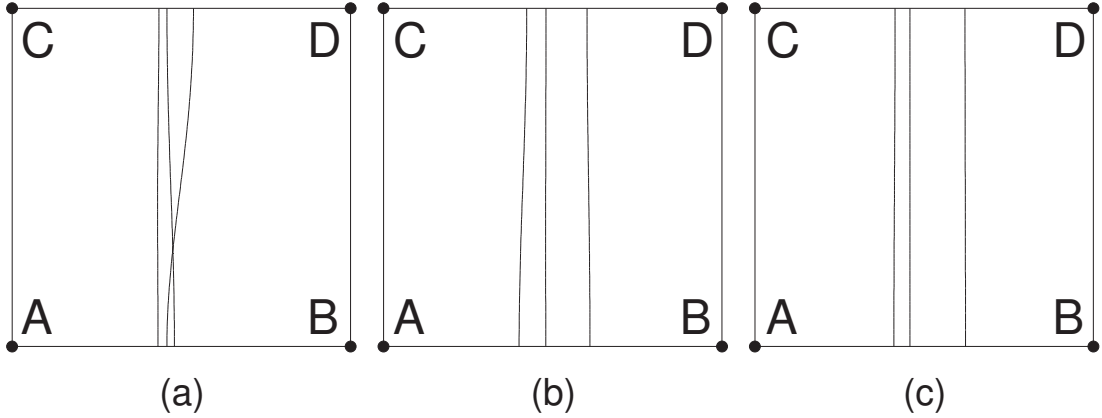}
\caption{Electronic $k_{z}$-dispersion for Sr$_{2}$RuO$_{4}$ along the
cut of the Fermi surface highlighted in \fref{Fig_EFSURF_xy}(a) by the
black bar. 27BM (a), M27BM1 (b) and M27BM2 (c). The corners of the
surfaces are $A = (0.2, 0.2,0) \frac{2 \pi}{a}$, $B = (0.45, 0.45, 0)
\frac{2\pi}{a}$, $C = \left(0.2 \frac{2\pi}{a}, 0.2 \frac{2\pi}{a},
\frac{2\pi}{c}\right)$, $D = \left(0.45 \frac{2\pi}{a}, 0.45 \frac{2
\pi}{a}, \frac{2 \pi}{c}\right)$.}\label{Fig_EFSURF_kz}
\end{figure}

The associated calculated phonon dispersion for the $\Lambda_{1}$ modes
of Sr$_{2}$RuO$_{4}$ along the $\Lambda \sim (0,0,1)$ direction is
depicted in \fref{Fig_SRO_LAM}(a) and for the relevant $\Delta_{1}$ and
$\Delta_{4}$ mode along the $\Delta \sim (1,0,0)$ direction in
\fref{Fig_SRO_DEL1_4}(a) together with the experimental data points
\cite{Braden07}. Figures \ref{Fig_SRO_LAM}(b), (c) and figures
\ref{Fig_SRO_DEL1_4}(b), (c), respectively, display our results for two
modified 27 band models (M27BM1 and M27BM2) with the tight-binding
parameters important for $c$-axis is coupling (O$_{xy}$-O$_{z}$, Ru-Sr,
O$_{xy}$-Sr, Ru-O$_{z}$) reduced by 1/2 and 1/5 with respect to the
LDA-based 27BM. This leads as shown in figures
\ref{Fig_EFSURF_kz}(a)-(c) to a much stronger anisotropy as in the
27BM, i.e. to a strongly reduced $k_{z}$-dispersion which is also much
weaker than in La$_{2}$CuO$_{4}$ \cite{Bauer09}.

\begin{figure}
\centering
\includegraphics[]{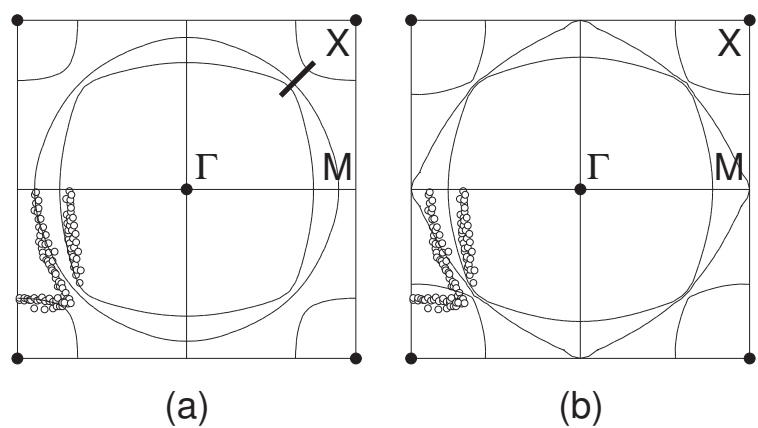}
\caption{Comparison of the calculated multi-sheet Fermi surface of
Sr$_{2}$RuO$_{4}$ in the $k_{z} = 0$ plane in model M27BM2 (a) and the
27BM (b) with the measured Fermi surface from \cite{Shen07}. The open
dots ($\circ$) give the experimental results. For the explanation of
the small black bar in (a) see
\fref{Fig_EFSURF_kz}.}\label{Fig_EFSURF_xy}
\end{figure}

\begin{figure}
\centering
\includegraphics[]{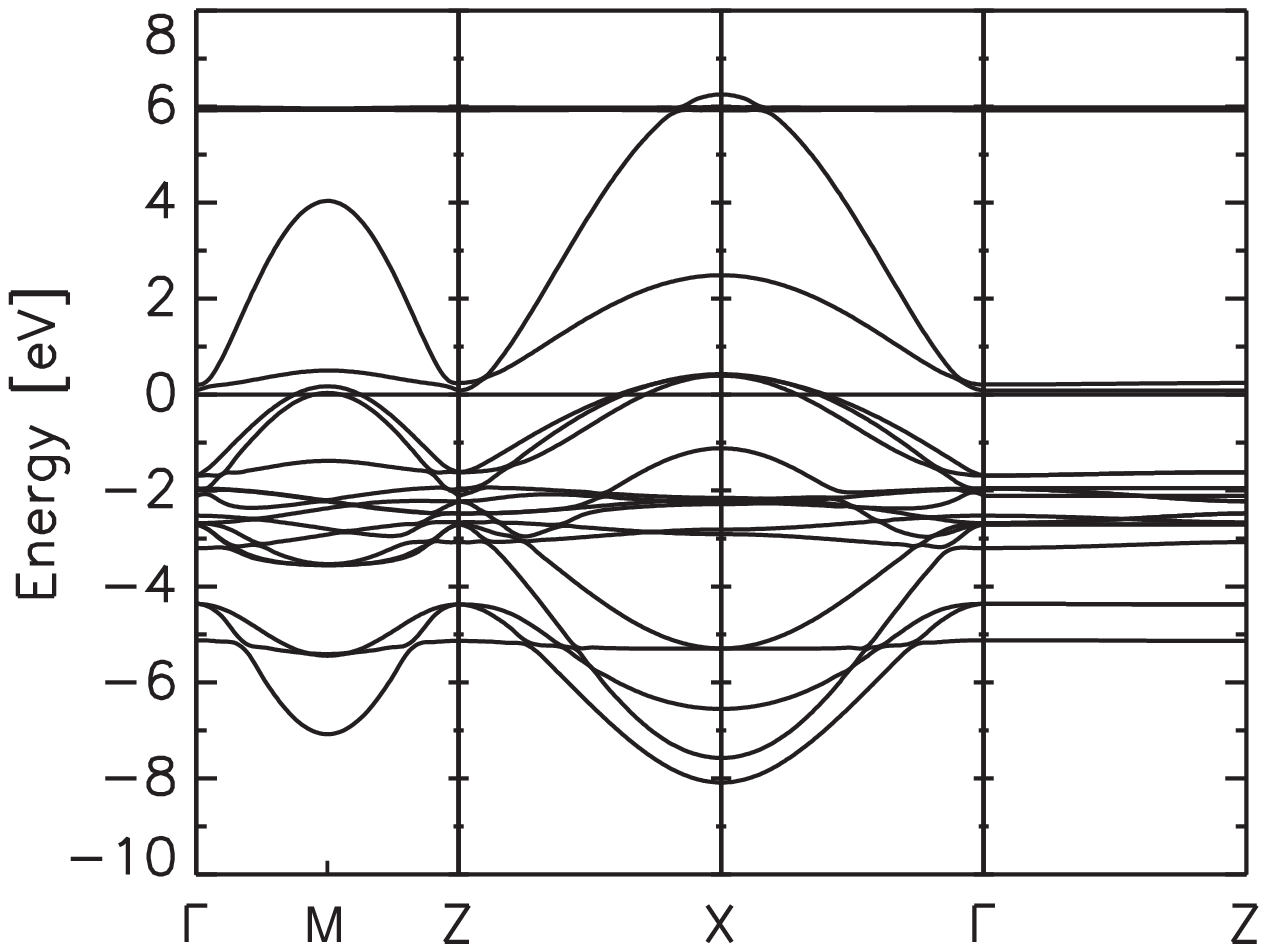}
\caption{Electronic bandstructure $E_{n} (\vc{k})$ of Sr$_{2}$RuO$_{4}$
in the M27BM2 taking into account the strongly enhanced anisotropy of
the real material as compared with the 27BM as a typical DFT-LDA based
model.}\label{Fig_SRO_EDP}
\end{figure}

In \fref{Fig_EFSURF_kz} the reduction of the electronic
$k_{z}$-dispersion is shown along the cut of the Fermi surface (FS)
highlighted in figure \ref{Fig_EFSURF_xy}(a) when going from the 27BM
to the most anisotropic M27BM2 via an intermediate $c$-axis coupling in
M27BM1. The best result for the phonon modes in figures
\ref{Fig_SRO_LAM} and \ref{Fig_SRO_DEL1_4}, which are less well
described in the 27BM, are obtained for the highly anisotropic M27BM2,
i.e. a nearly two-dimensional BS reflected by an even considerably
weaker $k_{z}$-dispersion as in case of La$_{2}$CuO$_{4}$, see
\fref{Fig_SRO_EDP} and \cite{Bauer09}.

It is enlightening to point out that in context with the $\Delta_{1}$
and $\Delta_{4}$ branches in \fref{Fig_SRO_DEL1_4} an investigation of
\cite{Braden07} exhibits that the dispersion cannot be described within
a normal ionic shell model extended by homogeneous electron gas
screening to simulate the metallic character of Sr$_{2}$RuO$_{4}$. Two
specially adapted force constants between Ru-O$_{z}$ and Ru-O$_{xy}$
had to be introduced to mimic the observed dispersion. Of course these
force constants do not have an intrinsic physical meaning. On the other
hand, from our calculation we can conclude that on a microscopic level
an accurate electronic dispersion along the $c$-axis is essential to
understand the observed mode behaviour.

\begin{figure}
\centering
\includegraphics[]{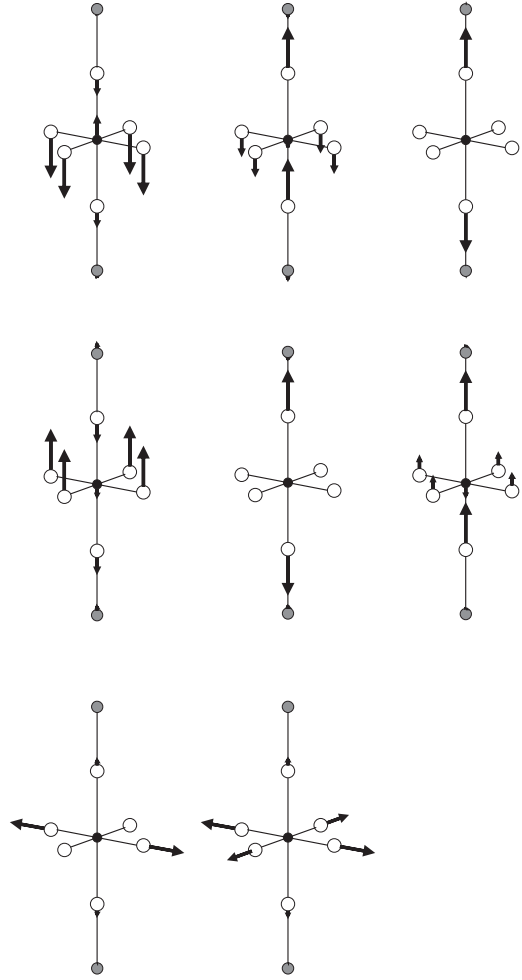}
\caption{Displacement patterns of certain phonon modes relevant for the
discussion of the phonon dynamics in Sr$_{2}$RuO$_{4}$. From left to
right we display in the first row $A^{\Gamma}_{2u}$(ferro),
$A^{\Gamma}_{2u} (\uparrow\downarrow)$, ${\rm O}^{\Gamma}_{z}$; in the
second row $A^{Z}_{2u} (\uparrow\downarrow)$, ${\rm O}^{Z}_{z}$,
$A^{Z}_{2u}$(ferro) and in the third row $\Delta_{1}/2$ and $O^{X}_{\rm
B}$.}\label{Fig_EVs}
\end{figure}

For a more detailed discussion of the rearrangement of the phonon modes
in figures \ref{Fig_SRO_LAM} and \ref{Fig_SRO_DEL1_4} as a result of a
modification of the $k_{z}$-dispersion of the BS it is useful to
display in \fref{Fig_EVs} the displacement patterns of some phonon
modes relevant for our studies.

In the 27BM the frequency of ${\rm O}^{Z}_{z}$ is nearly degenerate
with $A^{Z}_{2u}$(ferro). We denote the latter mode as "ferroelectric"
because its displacement pattern looks similar to that of the
ferro-like $A^{\Gamma}_{2u}$(ferro) mode at the $\Gamma$-point, see
\fref{Fig_EVs}. Here the oxygen anions vibrate coherently against the
cations in the lattice and as a consequence the electric dipole moments
generated by the motion add constructively. With decreasing
$k_{z}$-dispersion of the BS in M27BM1 and M27BM2 achieved by a
reduction of the relevant tight-binding parameters mentioned above by
1/2 and 1/5, respectively, the frequency of $A^{Z}_{2u}$(ferro)
increases by about 2 THz while ${\rm O}^{Z}_{z}$ remains virtually
unchanged. This highly sensible behaviour with regard to $c$-axis
coupling in Sr$_{2}$RuO$_{4}$ has also been demonstrated for
La$_{2}$CuO$_{4}$ \cite{Bauer09} where the frequency of $A^{Z}_{2u}$
(ferro) increases by more than 3 THz as can bee read off from
\fref{Fig_LCO_LAM}.

\begin{table}
\caption{Magnitudes of the charge fluctuations $|\delta
\zeta_{\kappa}|$ in units of $10^{-3}$ particles from equation
\eref{Eq11} excited on the Ru4d, O$_{xy}$2p, ${\rm O}_{z}$2p and Sr4d
orbitals in the ${\rm O}^{Z}_{z}$ and $A^{Z}_{2u}$(ferro) mode. The
results are given for comparison for the M27BM2 and the 27BM,
respectively.}\label{Tab_dzetakappa}
\begin{indented}
\lineup  \item[]\begin{tabular}{@{}*{9}{l}} \br & \centre{4}{O$_z^Z$}& \centre{4}{A$^Z_{2u}$(ferro)} \\
 &Ru4d & O$_{xy}$2p & O$_{z}$2p & Sr4d & Ru4d & O$_{xy}$2p & O$_{z}$2p & Sr4d \cr\mr
M27BM &26.98 & 7.62 & 0.02 & 0.49 & 0 & 0 & 3.22 & 0.34 \\ 27BM & 27.34
& 6.43 & 0.63 & 3.57 & 0 & 0 & 9.19 & 4.69
\end{tabular}
\end{indented}
\end{table}

The physical origin for the sensitivity derives from the fact that for
symmetry reasons in case of $A^{Z}_{2u}$(ferro) CF's can be excited in
the screening process only at the ions in the ionic layers, i.e.
O$_{z}$, La in La$_{2}$CuO$_{4}$ and O$_{z}$, Sr in Sr$_{2}$RuO$_{4}$.
The strength of these CF's on the other hand is governed by the matrix
elements of the proper polarization part $\Pi_{\kappa\kappa'}$ in
equation \eref{Eq10} of the corresponding out-of-plane ions and, thus,
depends critically on the coupling along the $c$-axis. Contrarily, the
renormalization of ${\rm O}^{Z}_{z}$ is nearly exclusively given by the
CF's on the Ru4d and O$_{xy}$2p orbitals in the plane. For quantitative
results compare with the CF's listed in \tref{Tab_dzetakappa}. This
explains the inertness of ${\rm O}^{Z}_{z}$ and the sensibility of
$A^{Z}_{2u}$(ferro) with respect to the change of the electronic
$k_{z}$-dispersion.

As a further example for the critical interrelation of $c$-axis phonon
dynamics and electronic $c$-axis coupling we note that the steep branch
connecting $A^{\Gamma}_{2u}$(ferro) with ${\rm O}^{Z}_{z}$ in
La$_{2}$CuO$_{4}$ is missing in Sr$_{2}$RuO$_{4}$. In the latter case
due to mode rearrangement $A^{\Gamma}_{2u}$(ferro) as the lowest of the
three highest modes at $\Gamma$ is connected to
$A_{2u}^{Z}\,(\uparrow\downarrow)$ that is the lowest of the three
highest modes at $Z$ and not to ${\rm O}^{Z}_{z}$ being the second
highest mode.

The BS underlying the M27BM2 is illustrated in \fref{Fig_SRO_EDP}. As
already mentioned the dispersion of the bands along the $\Lambda$
($k_{z}$) direction is extremely small, considerably smaller than for
the 27BM and for the case of La$_{2}$CuO$_{4}$. Thus Sr$_{2}$RuO$_{4}$
is a nearly two-dimensional Fermi liquid. However, as our calculation
in figures \ref{Fig_SRO_LAM} and \ref{Fig_SRO_DEL1_4} have shown the
remaining weak three-dimensionality is crucial to obtain a reliable
description of certain $c$-axis phonons and the $c$-axis charge
response in general.

Strong hybridization between Ru4d and O$_{xy}$2p states is evident from
the BS in \fref{Fig_SRO_EDP} because of the striking in-plane
dispersion of the bands. Moreover, we find a similar shape of the
partial density of states (PDOS) of the Ru4d and the O$_{xy}$2p
orbitals around the Fermi energy $\varepsilon_{F}$. The states at
$\varepsilon_{F}$ are more strongly of Ru4d type with some admixture of
O$_{xy}$2p. The contribution of O$_{z}$2p and Sr4d around
$\varepsilon_{F}$ is very small. In the M27BM2 we obtain for the PDOS
at $\varepsilon_{F}$ $Z_{{\rm Ru4d}}\,(\varepsilon_{F}) = 3.375$
eV$^{-1}$ , $Z_{{\rm O}_{xy}2p}\,(\varepsilon_{F}) = 0.795$ eV$^{-1}$,
$Z_{{\rm O}_{z}2p} (\varepsilon_{F}) = 0.002$ eV$^{-1}$ and $Z_{{\rm
Sr4d}} (\varepsilon_{F}) = 0.001$ eV$^{-1}$.

There are qualitative differences concerning the origin of
hybridization comparing Sr$_{2}$RuO$_{4}$ with La$_{2}$CuO$_{4}$.
Hybridization in Sr$_{2}$RuO$_{4}$ is strongly favoured because the
radial extent of the 4d wave functions in Ru is much larger than for
the 3d wave functions in Cu. Concurrently this leads to a weaker
influence of electron correlation effects, e.g. a smaller on-site
Coulomb repulsion $U$ for Ru4d compared to Cu3d. In our calculations we
find for $U$(Cu3d) = 1.005 dRy and for $U$(Ru4d) = 0.619 dRy. On the
other hand, hybridization is promoted in La$_{2}$CuO$_{4}$ as compared
to Sr$_{2}$RuO$_{4}$ because the energy levels of the copper and oxygen
ions are close by in La$_{2}$CuO$_{4}$ but far off between ruthenium
and oxygen in Sr$_{2}$RuO$_{4}$.

A further remarkable difference of the BS in Sr$_{2}$RuO$_{4}$ and
La$_{2}$CuO$_{4}$ can be read off from \fref{Fig_SRO_EDP} and the BS
for La$_{2}$CuO$_{4}$ in \cite{Bauer09}. In Sr$_{2}$RuO$_{4}$ up to
three bands close to each other are crossing the Fermi level giving
rise to a three-sheet FS while in La$_{2}$CuO$_{4}$ only one band is
crossing and other bands are more far away. Thus we have a completely
different situation as \textit{interband transitions} are concerned
which are on a much lower energy scale in Sr$_{2}$RuO$_{4}$ as in
La$_{2}$CuO$_{4}$. So we find for $E_{n'}(\vc{k} + \vc{q}) -
E_{n}(\vc{k})$ with $n' \not= n$ for $\vc{q}$ at the $Z$ point in case
of Sr$_{2}$RuO$_{4}$ a minimum of 20.67 meV in the M27BM2 around
$\vc{k} = 0.68 \frac{\pi}{a} (1,1,0)$, see also \fref{Fig_SRO_EDP},
while in La$_{2}$CuO$_{4}$ we obtain for the $Z$ point at $\vc{k} =
0.42 \frac{\pi}{a}(1,1,0)$ a minimum of 452.71 meV. This, of course,
has important consequences for the contribution of the interband
transitions to $\Pi_{\kappa\kappa'}$ in equation \eref{Eq10} concerning
magnitude as well as energy scale.

For example the low energy scale of the interband transitions in
Sr$_{2}$RuO$_{4}$ can generate damping of possible plasmons already at
very low energy. Such a damping already present in the collisionless
regime together with additional damping due to interactions between the
electrons very likely leads to overdamping of the very low lying
plasmons in Sr$_{2}$RuO$_{4}$ along the $c$-axis. The energy of the
latter in the collisionless regime is only 5-6 meV along the $\Lambda$
direction as will be shown in \sref{SecThreeThree}. Thus, the degree of
the remaining weak three-dimensionality quantified by our computations
is also very important for a possible existence of plasmons and coupled
phonon-plasmon modes around the $c$-axis. Such modes have been
predicted for La$_{2}$CuO$_{4}$ in our recent calculations
\cite{Bauer09} which, however, is not so anisotropic as
Sr$_{2}$RuO$_{4}$ and the plasmons along the $c$-axis are at higher
frequencies that overdamping can be avoided.

In figure \ref{Fig_EFSURF_xy}(a) and (b), respectively, the calculated
FS in the $k_{z} = 0$ plane is shown for the M27BM2 and the 27BM,
together with the experimental data points \cite{Shen07}. The FS
consists of the sheets $\alpha$, $\beta$, $\gamma$. The $\alpha$ sheet
is holelike and $\beta$ and $\gamma$ are electronlike. The $\gamma$
sheet is dominantly derived from the Ru4d$_{xy}$ orbitals, the $\alpha$
and $\beta$ sheets are primarily related to the 4d$_{xz}$ and 4d$_{yz}$
orbitals of Ru which exhibit anticrossing behaviour along the zone
diagonal. The large electronlike $\gamma$ sheet passes right through
the hybridization gap between the $\alpha$ and $\beta$ sheets.

The experimental results are not well described by the 27BM in
\fref{Fig_EFSURF_xy}(b). In particular the crossing between the $\beta$
and $\gamma$ sheet in the 27BM which is typical forLDA calculations is
experimentally not reproduced. This is really a crossing and not an
anticrossing as for $k_{z} = 0$ the bands related to d$_{xy}$ and
d$_{xz}$/d$_{yz}$ oribitals have different symmetry and no mixing is
allowed. On the other hand, our modified BS model M27BM2 in
\fref{Fig_EFSURF_xy}(a) is in good agreement with the experimental FS.

In order to obtain also a more global impression of the magnitude of
the enhanced anisotropy in the M27BM2 as compared to the 27BM we
compare some FS parameters being important for transport like the Drude
plasma energy tensor and the Fermi velocity tensor. The Drude tensor
ist defined as
\begin{equation} \label{Eq15}
\hbar \Omega_{p,ij} = \left( \frac{8\pi}{N V_{z}} \sum\limits_{\vc{k}n}
\delta (E_{n}(\vc{k}) - E_{F}) v_{\vc{k}n,i}
v_{\vc{k}n,j}\right)^{\frac{1}{2}}
\end{equation}
and the Fermi velocity is given by
\begin{equation} \label{Eq16}
\langle v_{F,ij}^{2}\rangle^{1/2} = \left(\frac{2}{N}
\sum\limits_{\vc{k}n} \theta (E_{n} (\vc{k}) - E_{F}) v_{\vc{k}n,i}
v_{\vc{k}n, j}\right)^{\frac{1}{2}},
\end{equation}
with
\begin{equation} \label{Eq17}
\vc{v}_{\vc{k}n} = \frac{1}{\hbar}
\frac{\partial\,E_{n}(\vc{k})}{\partial \vc{k}} .
\end{equation}

\begin{table}
\caption{Calculated data for the Fermi surface parameters (Drude plasma
energy tensor, Fermi velocity tensor) and the anisotropy ratio
$A_{\Omega} = \Omega_{p,xx}/\Omega_{p,zz}$; $A_{v_{F}} = \langle
v_{F,xx}^{2}\rangle^{1/2}/\langle v^{2}_{F,zz}\rangle^{1/2}$ in the
27BM and M27BM2 for Sr$_{2}$RuO$_{4}$. For comparison the corresponding
results are also given for La$_{2}$CuO$_{4}$
\cite{Bauer09}.}\label{Tab_aniso}

\begin{indented}
\lineup \item[]\begin{tabular}{@{}*{7}{l}} \br & $\hbar\Omega_{{\rm
p},xx}$ & $\hbar\Omega_{{\rm p},zz}$ & $\langle v^2_{{\rm
F},xx}\rangle^{1/2}$ & $\langle v^2_{{\rm F},zz}\rangle^{1/2}$ &
$A_\Omega$ & $A_{v_{\rm F}}$ \cr\mr 27BM & 1045.94 & 65.50 & 6.63 &
0.88 & 15.97 & 7.53 \\ M27BM2 & 1017.40 & 3.57 & 6.03 & 0.08 & 284.99 &
75.38 \\ M31BM & 648.60 & 25.25 & 2.99 & 0.11 & 25.69 & 27.18
\end{tabular}
\end{indented}
\end{table}

The output of our calculation for the 27BM and the M27BM2 is given in
\tref{Tab_aniso}. As can be seen we obtain a large enhancement of the
anisotropy ratio for the plasma frequencies $A_{\Omega}$ and of the
Fermi velocities $A_{v_{\rm F}}$ by about a factor 18 and 10,
respectively, in the M27BM2 as compared to the LDA-based 27BM.

Additionally, we have included in \tref{Tab_aniso} the corresponding
calculated data for La$_{2}$CuO$_{4}$ \cite{Bauer09}. Again we
recognize the by far larger anisotropy of Sr$_{2}$RuO$_{4}$ also in the
transport properties. The values for $\Omega_{p,zz}$ and $\langle
v_{F,zz}^{2}\rangle^{1/2}$ are significantly increased in
La$_{2}$CuO$_{4}$.

\subsection{Phonondynamics in Sr$_{2}$RuO$_{4}$ - Comparison with La$_{2}$CuO$_{4}$}\label{SecThreeTwo}

In this section we investigate the phonon dynamics of Sr$_{2}$RuO$_{4}$
in the main symmetry directions $\Delta \sim (1,0,0)$, $\Sigma \sim
(1,1,0)$ and $\Lambda \sim (0,0,1)$ and continue our discussion from
\sref{SecThreeOne} of a comparison of important characteristica as
found in our earlier calculations for La$_{2}$CuO$_{4}$.

For a definite investigation of the phonon renormalization induced by
the {\em nonlocal} EPI effects of DF and CF type mediated by the second
term in equation \eref{Eq4} a quantitative reference model for the
calculation of the phonon dispersion not including the nonlocal
screening effects but representing the important ionic component of
binding in the material is needed. Such a model sketched in
\sref{SecTwo} including approximately the {\em local} EPI effects is
provided by the ab initio RIM extended via covalent ion softening and
scaling of certain short-ranged pair potentials.

\begin{figure}
\centering
\includegraphics[angle=90,width=\linewidth]{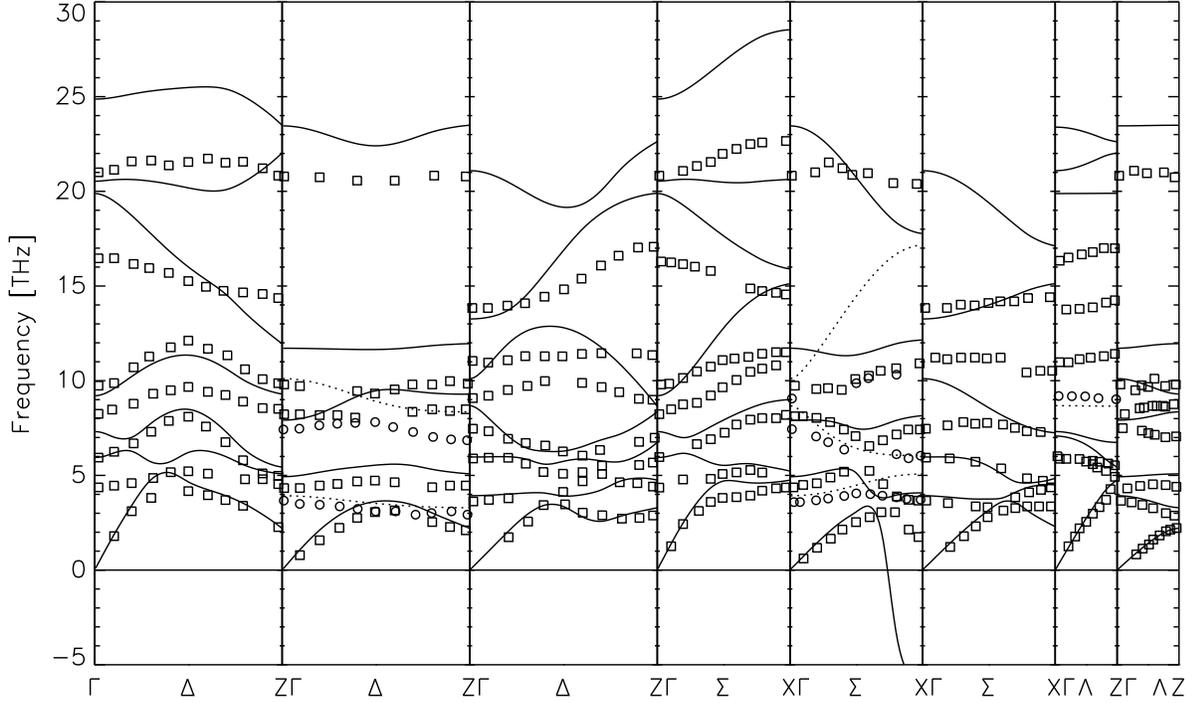}
\caption{Calculated phonon dispersion of Sr$_{2}$RuO$_{4}$ for the RIM
in the main symmetry directions $\Delta \sim (1,0,0)$, $\Sigma \sim
(1,1,0)$ and $\Lambda \sim (0,0,1)$. The various symbols representing
the experimental results are from \cite{Braden07} and indicate
different irreducible representations. The arrangement of the panels
from left to right according to the different irreducible
representations is as follows: $|\Delta_{1}|\Delta_{2} (\cdots,
\circ)$, $\Delta_{4} (-\!\!-, \Box)|\Delta_{3}|\Sigma_{1}|\Sigma_{2}
(\cdots, \circ)$, $\Sigma_{4} (-\!\!-, \Box)|\Sigma_{3}|\Lambda_{1}
(-\!\!-, \Box)$, $\Lambda_{2} (\cdots,
\circ)|\Lambda_{3}|$.}\label{Fig_SRO_PDP_RIM}
\end{figure}

The result of the phonon dispersion along the main symmetry directions
is shown for the RIM in \fref{Fig_SRO_PDP_RIM} and compared with the
experimental data from inelastic neutron scattering (INS)
\cite{Braden07}. The static effective charges found for the model are
Ru$^{2.7+}$, O$_{xy}^{-1.58}$, O$^{-1.67}_{z}$ and Sr$^{1.9+}$. In
particular the charges of Ru and O$_{xy}$ differ considerably from
their nominal values Ru$^{4+}$ and O$^{2-}$. This is due to the strong
hybridization of the Ru4d and O$_{xy}$2p states that reduces the
amplitude of the static effective charges in a mixed ionic-covalent
compound like Sr$_{2}$RuO$_{4}$ because of the charge transfer (CT)
from the cations to the anions is not complete as in the entirely ionic
case. Consistent with the result of the effective ionic charges is an
enhanced covalent character of the Ru-O$_{xy}$ plane with a dominant
covalent scaling of the Ru-O$_{xy}$ potential and a more ionic
character of the Sr-O$_{z}$ layer with an ionic scaling of the
Sr-O$_{z}$ potential as found in the calculations.

Altogether the RIM with these modifications yields good structural data
for the energy-minimized configuration. In detail we get for the planar
lattice constant $a = 3.834 \AA$, for the lattice constant along the
$c$-axis $c = 12.625 \AA$ and for the internal position of the O$_{z}$
and the Sr ion $z({\rm O}_{z})=0.163c$ and $z({\rm Sr})= 0.142c$. The
experimental values at 15 K are $a = 3.862 \AA$, $c = 12.723 \AA$,
$z({\rm O}_{z})=0.162c$, $z({\rm Sr})= 0.147c$ \cite{Chmaissem98}. From
these data we extract that the tetragonal distortion of the RuO$_{6}$
octahedra is significantly smaller than for the CuO$_{6}$ octahedra in
La$_{2}$CuO$_{4}$. The ratio of the in-plane to out-of-plane distance
of the oxygen is 0.9196 in Sr$_{2}$RuO$_{4}$ but 0.7706 in
La$_{2}$CuO$_{4}$.

The phonon dispersion of the RIM is in reasonable agreement with the
experiment as far as the branches with lower frequencies are concerned.
This points to the importance of the ionic component of binding in this
material. Large deviations are observed for the modes with higher
frequencies. In particular the high-frequency OBSM $\Delta_{1}/2$,
O$^{X}_{\rm B}$ and ${\rm O}^{Z}_{z}$, see figure \ref{Fig_EVs}, are
not well described. The corresponding frequencies are overestimated as
compared with the full calculation including DF's and CF's by about 5.8
THz, 7.4 THz and 6.7 THz, respectively. This can be attributed
according to our approach to the missing screening of the Coulomb
interaction by DF's and most importantly by CF's.

Noticeable is the soft $\Sigma_{4}$ mode at the $X$ point and the very
low frequency of the branch in the measurements (the notation for the
irreducible representations characterizing the symmetry of the modes
$\Sigma_{4}$ and $\Sigma_{3}$ is interchanged with the notation of
\cite{Braden07}). This mode is associated with the rotation of
RuO$_{6}$ octahedra around the $c$-axis. As indicated by our
calculations the softness of the rotational mode can be considered as a
precursor of a structural phase transition which in the meanwhile has
been observed in the Ca$_{2-x}$Sr$_{x}$RuO$_{4}$ series
\cite{Braden07}. From our calculation in the RIM this transition can be
expected to be driven essentially by the strong component of the ionic
forces in Sr$_{2}$RuO$_{4}$. Contrarily, Sr$_{2}$RuO$_{4}$ does not
experience the tilt instability related to the $\Sigma_{3}$ branch at
the $X$ point found in our calculations for La$_{2}$CuO$_{4}$. This
instability correctly indicates the experimentally observed structural
phase transition from the high-temperature tetragonal (HTT) to the
low-temperature orthorhombic (LTO) structure. Also this transition is
mainly brought about by the long-ranged ionic forces.

Substituting the smaller isovalent Ca ions for the Sr ions induces a
misfit between the Ca and Ru ions and generates a close connection
between the rotation of the RuO$_{6}$ octahedra and the electronic and
magnetic properties in the Ca$_{2-x}$Sr$_{x}$ RuO$_{4}$ series
\cite{Braden07}. Besides the softening of the rotational mode in the
range $0.5 \leq x \leq 1.5$ around $x \approx 0.5$ a similar structural
transition occurs due to the softening of the RuO$_{6}$ tilt mode
\cite{Moore08}. In this context it is interesting that at $x \approx
0.5$ in the paramagnetic phase of Ca$_{1.5}$Sr$_{0.5}$RuO$_{4}$ the
$\gamma$ Fermi sheet gains a remarkable $k_{z}$-dispersion and thus a
more three-dimensional character than Sr$_{2}$RuO$_{4}$ \cite{Uruma07}.
This issue also emphasizes an interrelation between the electronic
$k_{z}$-dispersion and the RuO$_{6}$ distortions.

\begin{figure}
\centering
\includegraphics[angle=90,width=\linewidth]{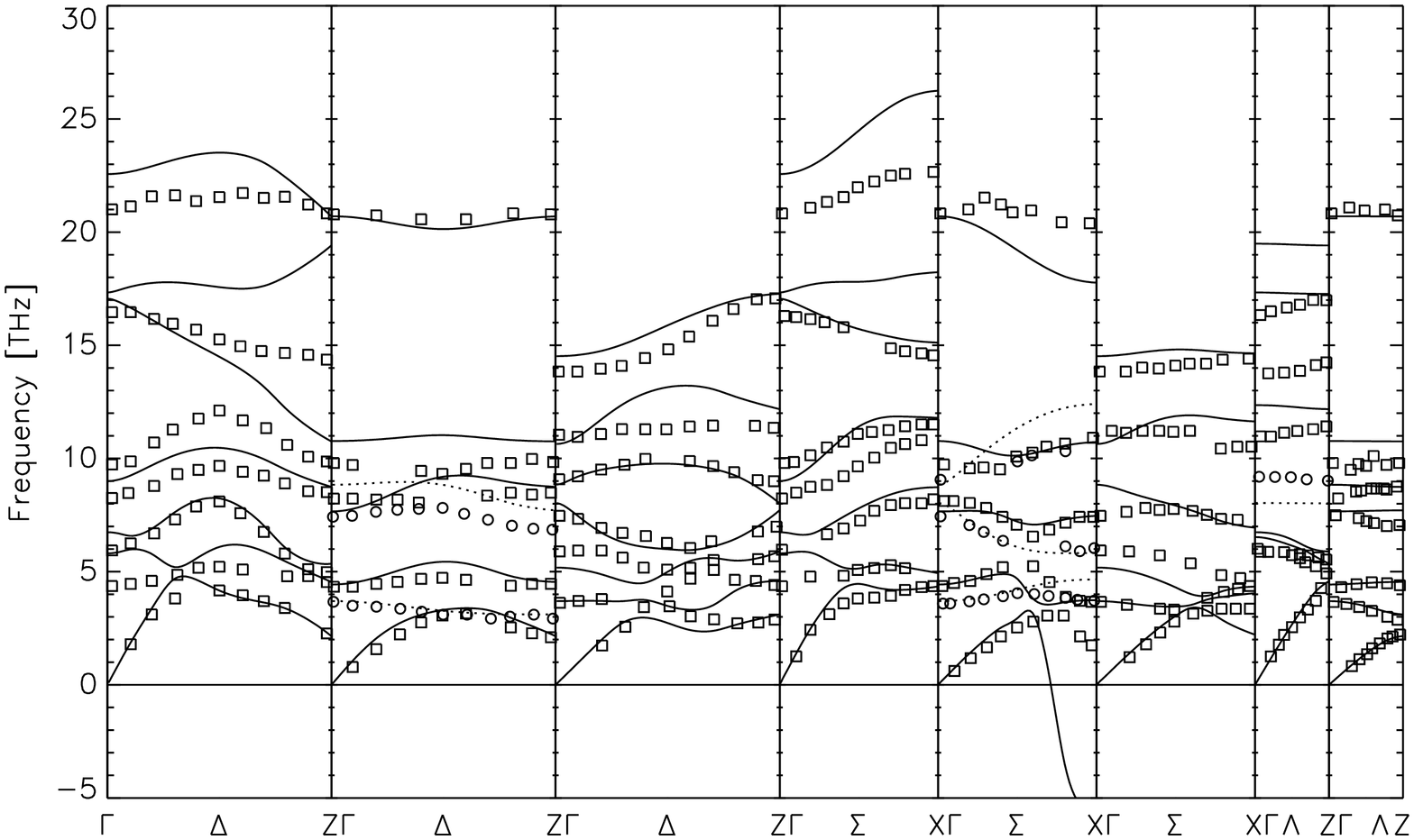}
\caption{Same as in \fref{Fig_SRO_PDP_RIM} but allowing additionally
dipole fluctuations in Sr$_{2}$RuO$_{4}$.}\label{Fig_SRO_PDP_D}
\end{figure}

In \fref{Fig_SRO_PDP_D} we display the calculated results of the phonon
dispersion admitting additionally to the RIM anisotropic DF's. This
leads to a better agreement of the dispersion curves with the inelastic
neutron results. Checking these calculations against the RIM the width
of the spectrum is reduced towards the experiment. The frequencies of
the transverse and longitudinal optical $E_{u}$ modes polarized in the
plane are decreased by the DF's and thereby the largest LO-TO splitting
is reduced at the $\Gamma$-point from 8.8 THz in the RIM to 6.3 THz
when dipolar polarization processes are included. At the same time the
frequencies of the $A_{2u}$ modes at $\Gamma$ and $Z$ are lowered
considerably by the DF's in $z$-direction and the large splitting of
the $A^{\Gamma}_{2u}$(ferro) mode of 10.13 THz in the RIM is reduced to
8.85 THz.

\begin{figure}
\centering
\includegraphics[angle=90,width=\linewidth]{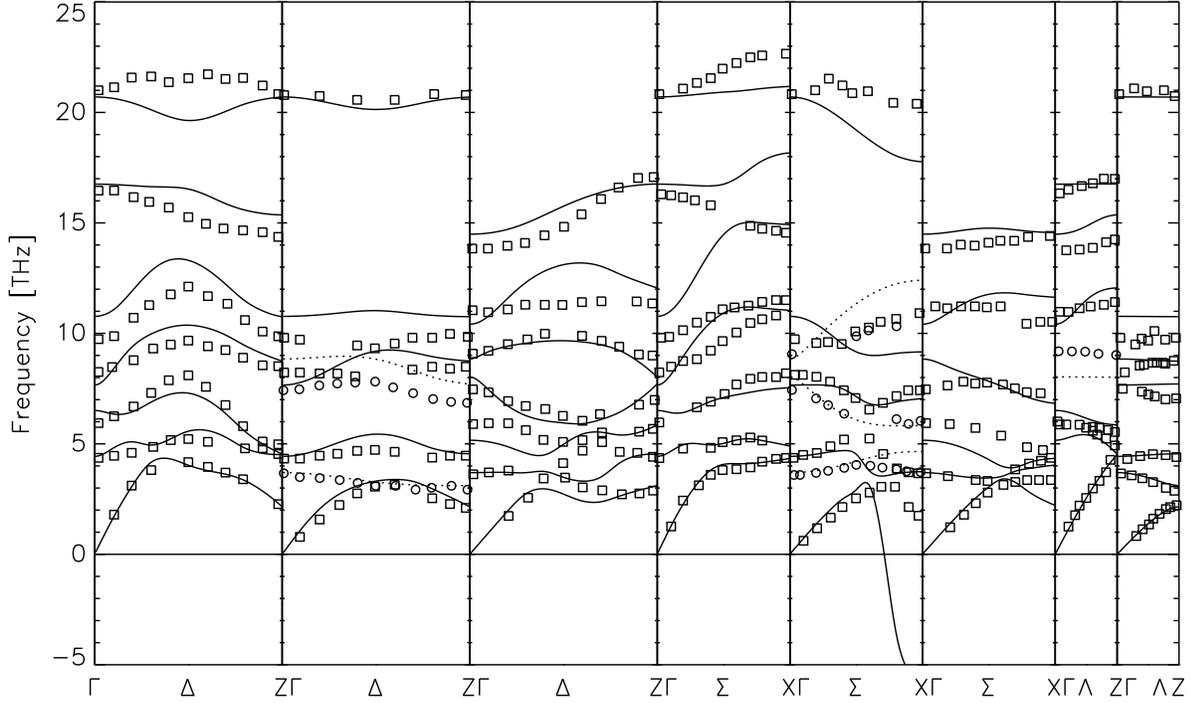}
\caption{Same as in \fref{Fig_SRO_PDP_RIM} but including additionally
dipole fluctuations and charge fluctuations on the basis of the M27BM2
in adiabatic approximation.}\label{Fig_SRO_PDP}
\end{figure}

The high-frequency OBSM $\Delta_{1}/2$, ${\rm O}^{X}_{\rm B}$ and ${\rm
O}^{Z}_{z}$ are also decreased due to the screening by the DF's but are
still overestimated by about 3.9 THz, 5.1 THz and 4 THz as compared
with the full calculation displayed in \fref{Fig_SRO_PDP} including
additionally CF's. Thus, the softening of these modes compared with the
RIM is of the order 2-3 THz and the remaining effect of the
renormalization can be assigned to the CF's on the Ru4d and O$_{xy}$2p
orbitals.

\begin{table}
\caption{Calculated charge fluctuations $\delta \zeta_{\kappa}$ in
Sr$_{2}$RuO$_{4}$ in the M27BM2 excited on the Ru4d, O$_{x}$2p,
O$_{y}$2p, O$_{z}$2p and Sr4d orbitals and induced by the OBSM
$\Delta_1/2$, ${\rm O}^{X}_{\rm B}$, ${\rm O}^{Z}_{z}$ and the ${\rm
O}^{\Gamma}_{z}$ mode. For the displacement patterns compare with
\fref{Fig_EVs}. Units are in $10^{-3}$ particles, a positive value
means an accumulation of electrons.}\label{Tab_dzetakappa_modes}

\begin{indented}
\lineup \item[]\begin{tabular}{@{}*{6}{l}} \br  & Ru4d & O$_x$2p &
O$_y$2p & O$_z$2p & Sr4d \cr\mr $\Delta_1/2$ & 38.252 & 0 & 4.472 &
-0.002 & 0 \\ O$^X_{\rm B}$ & 50.875 & 0 & 0 & 0.119 &
0.081 \\ O$^Z_z$ & 26.984 & 7.621 & 7.621 & 0.022 & 0.486 \\
O$_z^\Gamma$ & 12.521 & -3.714 & -3.714 & -2.224 & -0.319
\end{tabular}
\end{indented}
\end{table}

The quantitative results for the CF's according to equation \eref{Eq11}
are listed for the OBSM in \tref{Tab_dzetakappa_modes}. The CF's for
${\rm O}^{\Gamma}_{z}$ are also given in the table which prove to be
very different from those of ${\rm O}^{Z}_{z}$ despite the same local
displacement pattern. As can be extracted from \fref{Fig_EVs} for ${\rm
O}^{\Gamma}_{z}$ and ${\rm O}^{Z}_{z}$ the apex oxygens ${\rm O}_{z}$
move locally in phase against or away from the RuO layers. Because of
the weak screening along the $c$-axis and the corresponding large
strength of the nonlocal EPI we can expect these vibrations to induce
CF's in the RuO layers, see \tref{Tab_dzetakappa_modes} for the
calculated results. However, the CF's differ qualitatively between
${\rm O}^{Z}_{z}$ and ${\rm O}^{\Gamma}_{z}$, respectively, because for
the $\Gamma$-point vibrations the following sum rule can be derived for
the CF's $\delta\zeta_{\kappa}$ \cite{Falter93},
\begin{equation} \label{Eq18}
\sum\limits_{\kappa} \delta\zeta_{\kappa}\,(\Gamma\,\sigma) = 0\,\,.
\end{equation}

\begin{figure}
\centering
\includegraphics[width=\linewidth]{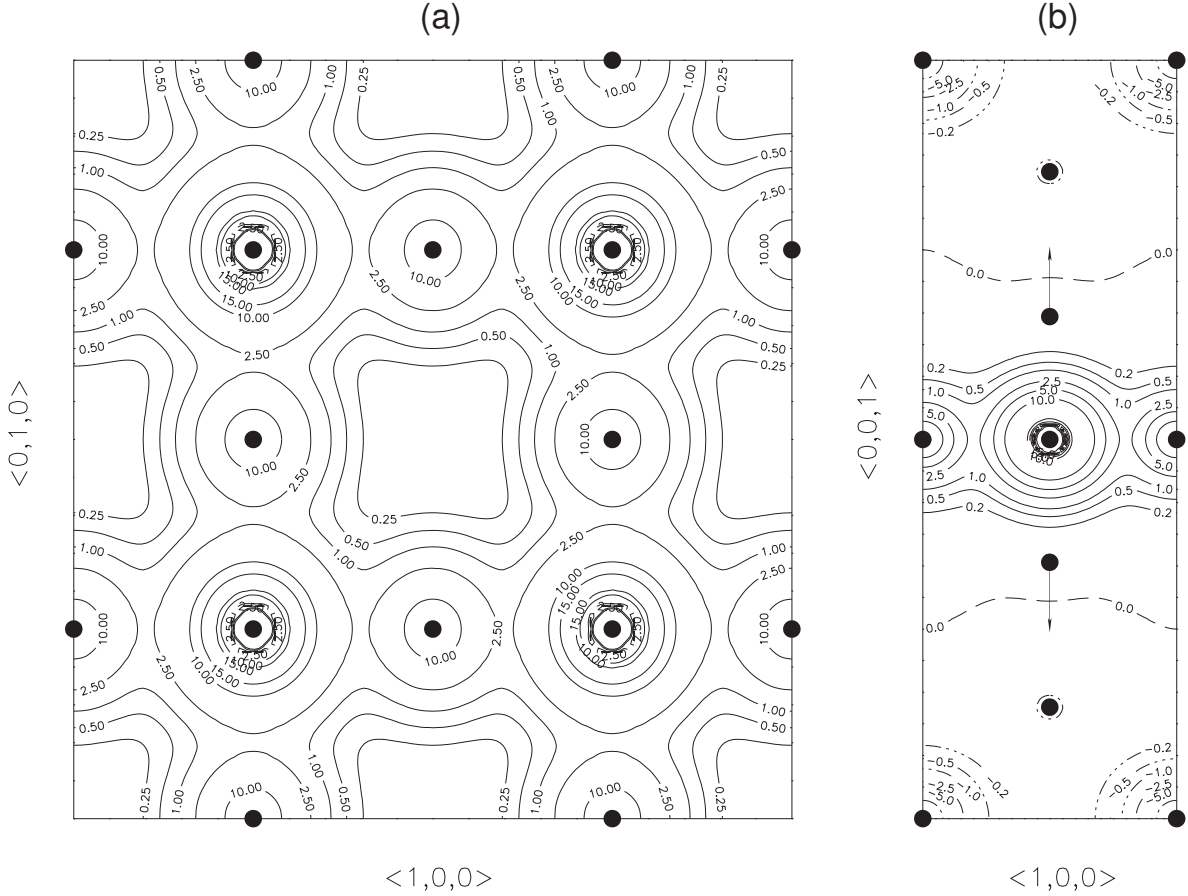}
\caption{Contour plot in the RuO plane (a) and perpendicular to the RuO
plane (b) of the nonlocal part of the displacement-induced charge
density redistribution for the ${\rm O}^{Z}_{z}$ mode calculated
according to equation \eref{Eq19} for the M27BM2. The units are in
$10^{-4} e^{2}/a_{\rm B}^{3}$. The phase of ${\rm O}^{Z}_{z}$ is as in
figure \ref{Fig_EVs}. Full lines ($-\!\!-$) indicate accumulation of
electrons in the corresponding region of space and broken lines
($-\cdots-$) indicate regions of repulsion for the
electrons.}\label{Fig_SRO_dRHO}
\end{figure}

The sum over $\kappa$ in equation \eref{Eq18} runs over the CF's in an
elementary cell of the crystal and $\sigma$ denotes the various
eigenmodes at $\Gamma$. Physically equation \eref{Eq18} means that
\textit{local charge neutrality} of the elementary cell is maintained
in the long-wavelength limit $(\vc{q} \to \vc{0})$, i.e. the mode
induced CT is organized within the cell such that from the outside the
cell looks electrically neutral. Equation \eref{Eq18} puts a strong
constraint on the possible CT in the cell and thus on the screening by
CF's. This restriction explains the small renormalization of the
frequency of only 0.581 THz of ${\rm O}^{\Gamma}_{z}$ in relation to
${\rm O}^{Z}_{z}$ where the renormalization due to CF's is 4.046 THz.
In \cite{Braden07} the pronounced softening and the broadening of ${\rm
O}^{Z}_{z}$ is attributed to a strong coupling between this phonon and
an interlayer charge transport. Such an interpretation is supported by
our calculations for ${\rm O}^{Z}_{z}$ where the CF's can be read off
from \tref{Tab_dzetakappa_modes} and the CT is illustrated in
\fref{Fig_SRO_dRHO} in terms of the nonlocal part of the displacement
induced rearrangement of the charge density according to
\begin{equation} \label{Eq19}
\delta \rho_{n} (\vc{r}, \vc{q} \sigma) = \sum\limits_{\vc{a},\kappa}
\delta \zeta^{\vc{a}}_{\kappa} (\vc{q},\sigma) \rho_{\kappa} (\vc{r} -
\vc{R}^{\vc{a}}_{\kappa}) .
\end{equation}
The CF's $\delta\zeta^{\vc{a}}_{\kappa}$ in equation \eref{Eq19} are
obtained from equation \eref{Eq11} and the form-factors from equation
\eref{Eq2}.

In contrast to the constraint for ${\rm O}^{\Gamma}_{z}$ expressed by
equation \eref{Eq18} no such restriction must be satistfied in the
metallic phase for ${\rm O}^{Z}_{z}$. While in ${\rm O}^{\Gamma}_{z}$
only an {\em intracell} CT is allowed summing up to zero, ${\rm
O}^{Z}_{z}$ generates CF's of the same sign in the cell
(\tref{Tab_dzetakappa_modes}). This finally leads to CF's of
alternating sign in consecutive RuO layers (cells), i.e. an {\em
interlayer (intercell)} CT is set up, see also \fref{Fig_SRO_dRHO}. The
latter provides an effective screening mechanism of the Coulomb
interaction and explains the strong renormalization of ${\rm
O}^{Z}_{z}$ as compared with ${\rm O}^{\Gamma}_{z}$.

In the adiabatic approximation used in the calculations so far the
interlayer CT is instantaneous. It has to be be replaced by a dynamic
collective charge transfer in case a phonon-plasmon scenario would be
realistic as for La$_{2}$CuO$_{4}$ in a nonadiabatic region nearby the
$c$-axis. The question if such a scenario is also likely for the more
anisotropic Sr$_{2}$RuO$_{4}$ is discussed in \sref{SecThreeThree}.

\begin{figure}
\centering
\includegraphics[]{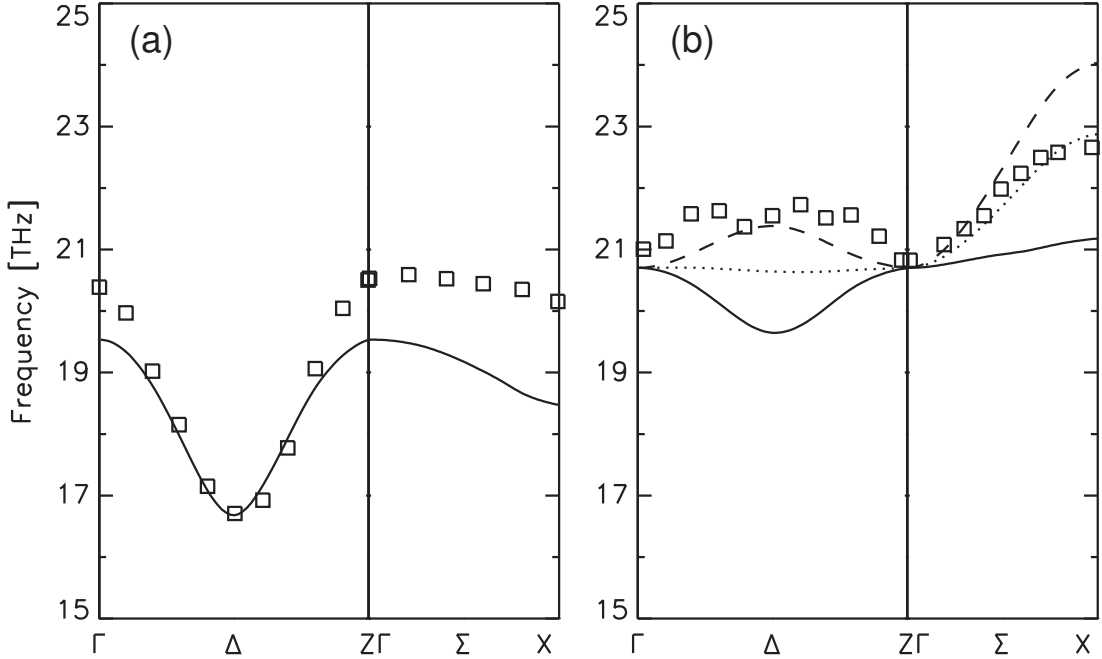}
\caption{Calculated dispersion of the highest $\Delta_{1}$ and
$\Sigma_{1}$ branch in La$_{2}$CuO$_{4}$ \cite{Falter06} (a) and
Sr$_{2}$RuO$_{4}$ (b). In (b) these branches have been obtained for
three different values of the on-site repulsion U(Ru4d) of the Ru4d
orbitals and are characterized by different line types, U(Ru4d)=0.619
dRy ($-\!\!-$), U(Ru4d)=0.819 dRy ($\cdots$) and U(Ru4d)=1.005 dRy
(-$\,$-$\,$-). The experimental values are indicated as open squares
$(\Box)$ \cite{Braden07}.}\label{Fig_PA_PDP}
\end{figure}

\begin{table}
\caption{ Calculated magnitudes of the charge fluctuations $|\delta
\zeta_{\kappa}|$ in units of $10^{-3}$ particles on the Ru4d orbitals
generated by the $\Delta_{1}/2$ and the ${\rm O}^{X}_{\rm B}$ mode in
dependence of the on-site Coulomb repulsion U(Ru4d). $\nu$ is the
frequency of the mode in units of THz. The results are listed from left
to right taking for U(Ru4d) 0.619 dRy, 0.819 dRy and 1.005 dRy. The
latter value matches the calculated value of U for the Cu3d orbital in
La$_{2}$CuO$_{4}$ and the first value is the calculated ab-initio
result for the Ru4d orbitals.}\label{Tab_modU}
\begin{indented}
\lineup \item[]\begin{tabular}{@{}*{7}{l}} \br &
\centre{3}{$\Delta_1/2$}& \centre{3}{O$^X_{\rm B}$} \cr\mr
$|\delta\zeta_{\rm Ru4d}|$ & 38.252 & 26.603 & 17.871 & 50.875 & 33.764
& 22.048 \\ $\nu$ & 19.635 & 20.642 & 21.381 & 21.178 & 22.875 & 24.031
\end{tabular}
\end{indented}
\end{table}

An important difference is found when considering the OBSM in the
cuprates and in Sr$_{2}$RuO$_{4}$, respectively. The anomalous
softening of the OBSM phononanomalies, see \fref{Fig_PA_PDP}(a), in
particular of the half breathing mode $\Delta_{1}/2$ which is typical
for La$_{2}$CuO$_{4}$ \cite{Falter06}, and most probably generic for
all the cuprate based HTSC's \cite{Pint05}, \cite{Graf08} is strongly
reduced in our calculation as compared with La$_{2}$CuO$_{4}$. In the
experiments the softening is completely absent, see
\fref{Fig_PA_PDP}(b). Our calculations show that there is a dominant
microscopic reason for the vanishing of the OBSM anomalies in
Sr$_{2}$RuO$_{4}$, namely the magnitude of the on-site Coulomb
repulsion U(Ru4d) of the 4d orbitals of ruthenium. The latter is
obviously underestimated in our ab-initio calculation based on the
ionic form-factor $\rho_{\kappa}$. Compared with the calculated value
for the on-site repulsion U(Cu3d)=1.005 dRy in La$_{2}$CuO$_{4}$ which
leads to good results for the frequencies of the OBSM in this material,
the calculated value for ruthenium, U(Ru4d) = 0.619 dRy, seems not so
reliable because of the more incomplete 4d shell of Ru and the
spherical average assumed in our computations for the orbital
densities. So, we increased the value of the Ru4d on-site repulsion in
two steps towards the calculated result for the Cu3d orbital for
comparison. As a consequence the phonon anomalies completely vanish in
these calculations in agreement with experiment as displayed in
\fref{Fig_PA_PDP}(b). Simultaneously by increasing U(Ru4d) the CF's on
the Ru4d orbitals that are a measure for the nonlocal electron-phonon
coupling are strongly decreased as can be seen in \tref{Tab_modU}.

\begin{figure}
\centering
\includegraphics[]{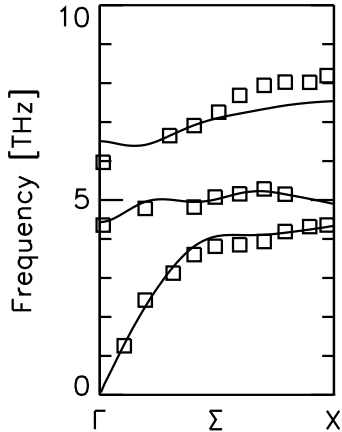}
\caption{Calculated phonon dispersion in the M27BM2 of the three lowest
$\Sigma_{1}$ branches displaying a complex anticrossing behaviour. The
open squares $(\Box)$ are the experimental data
\cite{Braden07}.}\label{Fig_SIG1}
\end{figure}

In \cite{Sidis99} and \cite{Braden02} sharply peaked magnetic
fluctuations at $\vc{q}_{0} = 0.6 \frac{\pi}{a} (1,1,0)$ have been
observed for Sr$_{2}$RuO$_{4}$. These fluctuations have been related to
dynamical nesting properties between the flat $\alpha$ and $\beta$
sheets of the FS, compare with \fref{Fig_EFSURF_xy}(a). Braden et al.
\cite{Braden07} have performed a search for Kohn anomalies in the
phonon spectrum related to this nesting structure. Promising candidates
would be the high frequency $\Sigma_{1}$ modes which are coupled to the
Ru4d CF's. However no signature for a Kohn anomaly is present in the
experiments as well as in the calculation in \fref{Fig_SRO_PDP}.
Another possibility discussed in \cite{Braden07} is the longitudinal
acoustic $\Sigma_{1}$ branch where a small dip is observed just around
$\vc{q}_{0}$, see \fref{Fig_SIG1}. From our calculations it seems very
unlikely that this dip is a nesting effect because our computations of
the phonon dispersion shows that this feature is well described by a
complex anticrossing between the three lowest branches with
$\Sigma_{1}$ symmetry, as shown in \fref{Fig_SIG1}. So we do not find
in Sr$_{2}$RuO$_{4}$ any evidence for EPI driven by nesting of the FS.

\subsection{Search for a phonon-plasmon scenario in Sr$_2$RuO$_4$}\label{SecThreeThree}

We investigate for Sr$_{2}$RuO$_{4}$ the possibility of a
phonon-plasmon scenario along the $c$-axis and in a small region around
this axis by performing calculations of the coupled electron-phonon
dynamics in the nonadiabatic regime. This is achieved approximatively
by allowing the proper polarization part $\Pi_{\kappa\kappa'}$ of the
DRF to depend on the frequency of the perturbation. Different from {\em
static screening} in the adiabatic approximation this leads to {\em
dynamical screening} of the Coulomb interaction. A nonadiabatic
treatment of the charge response and a related phonon-plasmon coupling
has been shown recently to be a realistic option for La$_{2}$CuO$_{4}$
if the real anisotropy of the material is taken into account in the
theory \cite{Bauer09}.

\begin{figure}
\centering
\includegraphics[]{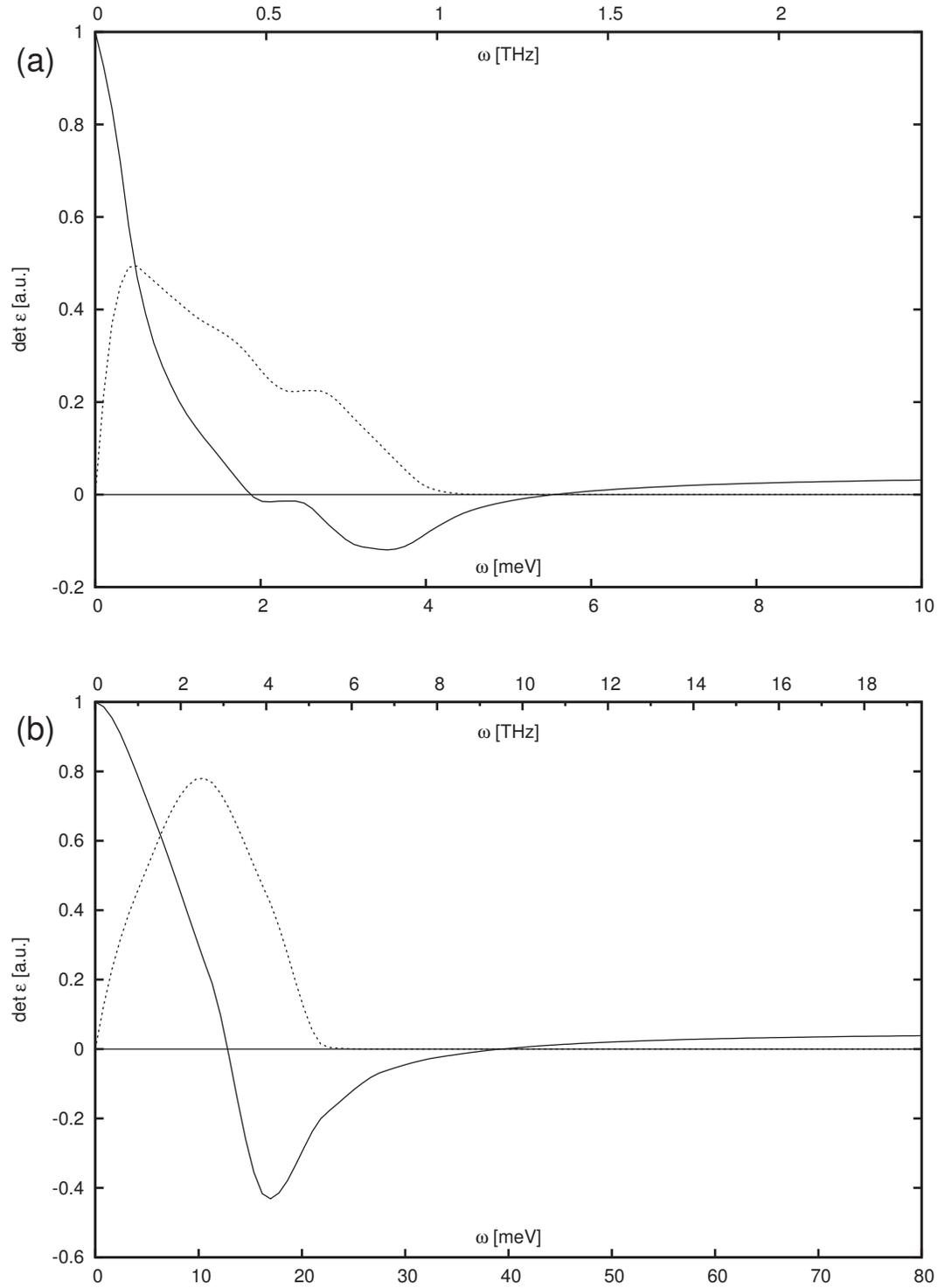}
\caption{Real- and imaginary part of $\det [\varepsilon_{\kappa\kappa'}
(\vc{q}, \omega)]$ at the $Z$ point in arbitrary units for
Sr$_{2}$RuO$_{4}$ in the M27BM2 (a) and for La$_{2}$CuO$_{4}$ in the
M31BM \cite{Bauer09} (b). The real part is given by the full line
($-\!\!-$) and the imaginary part by the dotted line
($\cdots$).}\label{Fig_deteps}
\end{figure}

Due to the much weaker $k_{z}$-dispersion of the electronic
band-structure obtained in the computation for Sr$_{2}$RuO$_{4}$ as
compared with La$_{2}$CuO$_{4}$ we can expect much lower plasmon
frequencies along the $c$-axis. This conjecture can be shown to be true
for example by calculating at the $Z$ point the free-plasmon frequency
according to equation \eref{Eq14} and checking the result against
La$_{2}$CuO$_{4}$, see \fref{Fig_deteps}. From the zero crossing of the
real and the imaginary part of $\det
[\varepsilon_{\kappa\kappa'}(Z,\omega)]$ we get for Sr$_{2}$RuO$_{4}$
in the M27BM2 the very small value of $\omega_{p} (Z) = 5.55$ meV
(1.342 THz) while for La$_{2}$CuO$_{4}$ we have $\omega_{p} (Z) =
39.33$ meV (9.509 THz) \cite{Bauer09} which is by a factor of seven
larger. This also expresses the much stronger anisotropy of
Sr$_{2}$RuO$_{4}$ in relation to the cuprates.

In the analysis of the phonon-plasmon scenario we have assumed the
collisionless regime, i.e. the quasiparticles (QP) do not scatter each
other and damping is only possible via intraband electron-hole decay
and low-lying interband transitions. However, in the cuprates and also
in Sr$_{2}$RuO$_{4}$ QP scattering due to electron-electron interaction
and other degrees of freedom is certainly important. This leads to
damping and possibly overdamping of $c$-axis plasmons at sufficiently
low frequencies and so the latter may cease to be well defined
collective excitations of the Fermi liquid.

In order to make an estimate of the range of frequencies where $c$-axis
plasmons should exist we examine the zero crossing of the real part of
the well known dielectric function of an electron gas
\begin{equation} \label{Eq20}
\varepsilon (\omega) = \varepsilon_{\infty} -
\frac{\omega_{p}^{2}}{\omega^{2} + \gamma^{2}} .
\end{equation}
$\varepsilon_{\infty}$ is the high-frequency dielectric constant,
$\gamma = \frac{1}{\tau}$ the scattering rate of the QP's along the
$c$-axis. Equating $\varepsilon (\omega)$ to zero we have the equation,
\begin{equation} \label{Eq21}
\omega^{2} = \frac{\omega^{2}_{p}}{\varepsilon_{\infty}} - \gamma^{2} .
\end{equation}
Using for $\gamma$ the value given in \cite{Tamasaku92} in the normal
state of La$_{2}$CuO$_{4}$ and the calculated value for the
high-frequency dielectric constant along the $c$-axis
$\varepsilon_{\infty}^{zz} \approx 2$ \cite{Falter02} we obtain
$\omega^{2} \geq 0$ for $\omega_{p} \geq 7.21$ THz. For smaller plasmon
frequencies we have $\omega^{2} < 0$, i.e. the plasmon has a diffusive
pole and hence the $c$-axis plasmon is overdamped in the normal state.
It should be remarked that larger values for
$\varepsilon_{\infty}^{zz}$ would increase the critical value for
$\omega_{p}$ beyond that the plasmon could exist. For example setting
$\varepsilon^{zz}_{\infty} = 3$ or $\varepsilon^{zz}_{\infty} = 4$ we
would have $\omega_{p} \geq 8.83$ THz or 10.19 THz, respectively.
Assuming a similar $c$-axis scattering rate of the QP's for
Sr$_{2}$RuO$_{4}$ we regard $\omega_{p} \approx 7$ THz as a reasonable
limiting frequency beyond which plasmons most likely are not
overdamped.

\begin{figure}
\centering
\includegraphics[]{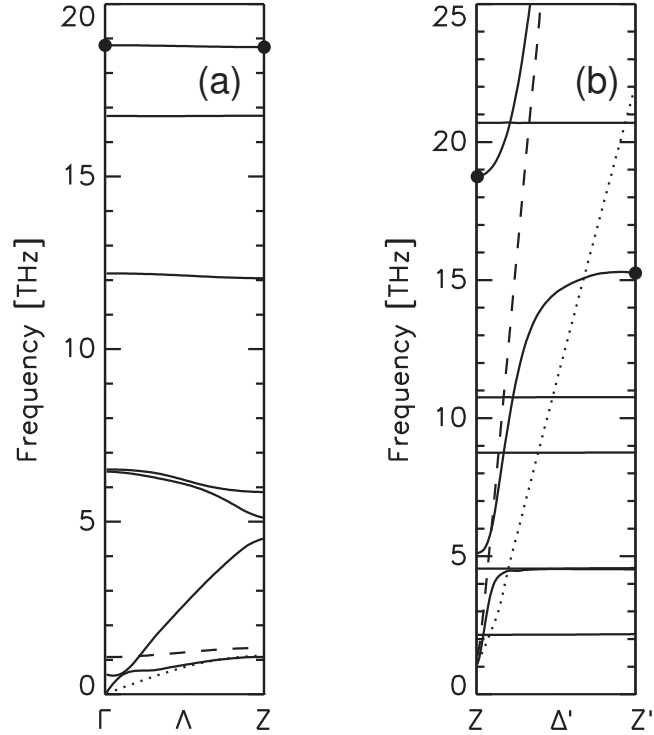}
\caption{Calculated nonadiabatic coupled phonon-plasmon dispersion
within the M27BM2. (a) $\Lambda_{1}$ modes ($-\!\!-$) in the
collisionless regime, the coupling modes at $\Gamma$
$(A^{\Gamma}_{2u}$(ferro, na) and at $Z ({\rm O}^{Z}_{z}$(na)) are
shown as black dots ($\bullet$). (b) $\Delta_{1}'$ modes ($-\!\!-$)
along the $\Delta' = \left(\varepsilon \frac{2\pi}{a},
0,\frac{2\pi}{c}\right)$ direction from $Z\,(\varepsilon = 0)$ to $Z'
(\varepsilon = 0.02)$. The black dot ($\bullet$) at $Z$ and $Z'$ are
the ${\rm O}^{Z}_{z}$(na) and approximately the ${\rm O}_{z}^{Z}$ mode
in adiabatic approximation, respectively. In both figures the broken
line (-$\,$-$\,$-) denotes the free-plasmon branch and the dotted line
($\cdots$) the borderline for damping due to electron-hole
decay.}\label{Fig_SRO_LAM_DEL_NA}
\end{figure}

Thus, from our calculations of the plasmon along the $c$-axis
($\Lambda$ direction) displayed in \fref{Fig_SRO_LAM_DEL_NA}(a) which
are at very low frequencies around 1 THz we conclude that the latter
are overdamped. On the other hand at larger plasmon frequencies which
result in case the wavevector $\vc{q}$ is not strictly parallel to the
$c$-axis, as shown for example in \fref{Fig_SRO_LAM_DEL_NA}(b), the
plasmon can acquire enough energy to resist overdamping and may exist
as a collective excitation.

In \fref{Fig_SRO_LAM_DEL_NA}(a) we identify instead of the six
$\Lambda_{1}$ branches of the adiabatic approximation (compare with
\fref{Fig_SRO_LAM}(c)) an additional branch due to phonon-plasmon
coupling. The free-plasmon has frequencies around 1 THz and is nearly
dispersionless along the $\Lambda$ direction. This is in contrast with
the situation in La$_{2}$CuO$_{4}$ \cite{Bauer09} where the
free-plasmon has significant higher frequencies and increases from
about 9.5 THz at the $Z$ point to about 12.8 THz at $\Gamma$. Thus, the
plasmon is not overdamped along the $c$-axis in La$_{2}$CuO$_{4}$ while
the phonon-plasmon scenario calculated in the collisionless regime for
Sr$_{2}$RuO$_{4}$ as displayed in \fref{Fig_SRO_LAM_DEL_NA}(a) is not
realistic.

\begin{figure}
\centering
\includegraphics[]{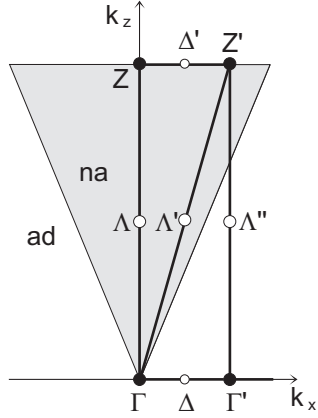}
\caption{Schematic representation of the nonadiabatic region in the
($k_{x}$, $k_{z}$) plane with the directions $\Delta' =
\left(\varepsilon \frac{2\pi}{a}, 0, \frac{2\pi}{c}\right)$, $\Lambda'
= \zeta \left(\varepsilon \frac{2\pi}{a}, 0, \frac{2\pi}{c}\right)$ and
$\Lambda'' = \left(\varepsilon \frac{2\pi}{a}, 0, \zeta
\frac{2\pi}{c}\right)$. $\zeta \in [0,1]$. na: nonadiabatic region; ad:
adiabatic region.}\label{Fig_cone}
\end{figure}

Analogous to the situation in La$_{2}$CuO$_{4}$ \cite{Falter05,Bauer09}
the calculated free-plasmon frequencies rapidly increase in proportion
to the transverse component of the $\vc{q}$ vector in a small
nonadiabatic region of the charge response around the $c$-axis. We have
performed calculations along the $\Lambda'' = \left(\varepsilon
\frac{2\pi}{a},0, \zeta \frac{2\pi}{c}\right)$ and $\Lambda' = \zeta
\left(\varepsilon \frac{2\,\pi}{a},0, \frac{2\pi}{c}\right)$ direction
in this region, see \fref{Fig_cone}, for different small
$\varepsilon$-values to study a possible phonon-plasmon scenario in
Sr$_{2}$RuO$_{4}$. We do not go into the details of these computations
here but what is important concerning the question of overdamping is
that for $\varepsilon \geq 0.0040$ in case of the $\Lambda'$ direction
and for $\varepsilon \geq 0.0025$ in the $\Lambda''$ direction the
free-plasmon frequency is larger than 7 THz, the estimated value for
the phonon-plasmon scenario to exist. From the calculations we find
that the longitudinal $A^{\Gamma}_{2u}$(ferro, na) mode at $\Gamma$
(\fref{Fig_SRO_LAM_DEL_NA}(a)) is virtually unchanged at $\varepsilon =
0.0040$ and so does exist around 19 THz. The same holds true for ${\rm
O}^{Z}_{z}$(na) at the $Z$ point being practically the same as ${\rm
O}^{Z'}_{z}$ for $\varepsilon = 0.0025$ in the $\Lambda''$ direction.
Similar as for La$_{2}$CuO$_{4}$ \cite{Bauer09} with increasing
$\varepsilon$ $A^{\Gamma}_{2u}$(ferro, na) and ${\rm O}^{Z'}_{z}$(na)
rapidly leave the spectrum to high frequencies.

In case of ${\rm O}^{Z'}_{z}$(na) this mode behaviour can also be
extracted from \fref{Fig_SRO_LAM_DEL_NA}(b) where we have presented our
calculated results of the dispersion of the coupled phonon-plasmon
$\Delta'_{1}$ modes along the $\Delta' = \left(\varepsilon
\frac{2\pi}{a},0,\frac{2\pi}{c}\right)$ direction from $Z =
\left(0,0,\frac{2\pi}{c}\right)$ to $Z' = \left(\varepsilon
\frac{2\pi}{a},0,\frac{2\pi}{c}\right)$ for $0 \leq \varepsilon \leq
0.02$. The highest mode at $Z$ is the ${\rm O}^{Z}_{z}$(na) mode also
seen in \fref{Fig_SRO_LAM_DEL_NA}(a) as a black dot. The broken line is
the dispersion of the free-plasmon branch calculated from equation
\eref{Eq14} and the dotted line is the borderline for damping due to
electron-hole decay investigated from ${{\rm max} \atop \vc{k} \in {\rm
BZ}}~(E_{n} (\vc{k}) - E_{n} (\vc{k} + \vc{q}))$ for the bands crossing
the Fermi level. From this figure the range of the region with a
nonadiabatic charge response can be estimated. It is characterized by
the steep branch which ultimately converges to the frequency of the
adiabatic ${\rm O}^{Z}_{z}$ mode (black dot at $Z'$). Similar as in
La$_2$CuO$_4$ this nonadiabatic region is very small and can be
estimated from the figure at about $\varepsilon = 0.01$. Unlike the
case of La$_{2}$CuO$_{4}$ the phonon-plasmon dispersion in
\fref{Fig_SRO_LAM_DEL_NA}(b) should not exist in the full range of
frequencies according to our estimate for overdamping below about 7
THz.

Parallel to the investigation of the ${\rm O}^{Z}_{z}$ mode in
La$_{2}$CuO$_{4}$ \cite{Falter05,Bauer09}, the line-broadening of this
mode observed by \cite{Braden07} in Sr$_{2}$RuO$_{4}$ can be understood
from the calculated phonon-plasmon scenario. Experimentally there is a
limited wave vector resolution in INS for the transverse direction
perpendicular to the $c$-axis which for the experiments in
La$_{2}$CuO$_{4}$ is on the average $\varepsilon = 0.03$ \cite{Pintpc}.
Thus, the relevant frequency range sampled in the measurement for ${\rm
O}^{Z}_{z}$ is over the steep branch beyond which the mode develops an
${\rm O}^{Z}_{z}$-like displacement pattern. According to the
calculations this sampling occurs for mode frequencies larger than 12.5
THz and leads to a corresponding broadening of the linewidth for ${\rm
O}^{Z}_{z}$ observed in the experiment.

Because the region with a metallic adiabatic charge response outweights
by a factor of roughly three the nonadiabatic region we can attribute
to the ${\rm O}^{Z}_{z}$ mode an adiabatic frequency of about 15.3 THz.
The $\Delta_{1}'$ branch starting at ${\rm O}^{Z}_{z}$(na) in
\fref{Fig_SRO_LAM_DEL_NA}(b) has not been observed in the experiments
so far because a very high $\vc{q}$-space solution transverse to the
$c$-axis would be needed to resolve the mode dispersion in the small
nonadiabatic sector.

In \cite{Bauer09} we have pointed out for La$_{2}$CuO$_{4}$ the
relevance of a strongly coupling nonadiabatic ${\rm O}^{Z}_{z}$ mode.
So it is interesting to see if such a large EPI also exists for ${\rm
O}^{Z}_{z}$(na) in Sr$_{2}$RuO$_{4}$. As a parameter for the strength
of the EPI in a certain mode $(\vc{q}\sigma)$ we have used in the past
the orbital averaged changes of the selfconsistent crystal potential
$\delta V_{\kappa} (\vec{q}\sigma)$, \cite{Falter05,Bauer09}. In case
of La$_{2}$CuO$_{4}$ we obtain in units of meV $|\delta V_{{\rm Cu3d}}|
= 955.98$ and $|\delta V_{{\rm O2p}}| = 515.98$ which should be
compared with the considerably reduced strength $|\delta V_{{\rm
Ru4d}}| = 277.24$ and $|\delta V_{{\rm O2p}}| = 316.64$ in
Sr$_{2}$RuO$_{4}$.

Altogether, we conclude from our calculations for Sr$_{2}$RuO$_{4}$
that nonadiabatic EPI via phonon-plasmon coupling is not possible
strictly along the $c$-axis but most likely away from this axis in a
very small region at higher free-plasmon frequencies. Moreover, the
strength is significantly smaller as in La$_{2}$CuO$_{4}$.

We have argued above that the longitudinal ferroelectric
$A^{\Gamma}_{2u}$(ferro, na) mode exists in the nonadiabatic sector
around the $\Gamma$ point at about 19 THz. This nonadiabatic result
helps to understand why $c$-axis optical activity seen in the
experiments \cite{Katsufuji96} is possible despite the fact that
Sr$_{2}$RuO$_{4}$ is in the metallic phase. In \cite{Katsufuji96} three
$c$-polarized $A_{2u}$ phonon structures at 6.04 THz, 10.88 THz and
14.51 THz have been detected which agree well with the calculated
transverse $A^{\Gamma}_{2u}$ modes in the M27BM2 at 5.16 THz , 10.42
THz and 14.49 THz. The by far largest oscillator strength is obtained
for the mode at 10.88 THz and according to our calculation this is the
ferroelectric mode $A^{\Gamma}_{2u}$(ferro). Consistent with the large
oscillator strength of this mode in the experiment our nonadiabatic
result for the longitudinal ferroelectric mode $A_{2u}^{\Gamma}$
(ferro, na) predicts a very large $A_{2u}$ splitting of about 9 THz.
This splitting dominates the infrared response for polarization along
the $c$-axis. Further we find in the calculations two other
longitudinal $A_{2u}$ modes at 12.379 THz and 6.964 THz which define
corresponding splittings with the transverse modes above.

From a theoretical point of view it is important to remark that optical
activity in the metallic phase cannot be explained using the adiabatic
approximation for a calculation of the phonon dispersion as is commonly
done by applying static DFT for the metal
\cite{Savrasov96,Wang99,Bohnen03,Giustino08}. In such calculations
there will be no LO-TO or $A_{2u}$ splittings being a measure for the
oscillator strength because the transverse effective charges vanish if
the metallic phase is treated in adiabatic approximation and
consequently the induced dipole moments defining the oscillator
strength in the dielectric function (matrix) vanish too. Thus, there
will be no optical activity in a metal by the phonons treated in
adiabatic approximation which underlies static DFT.

\section{Summary and conclusions} \label{SecFour}
We have shown that for a reliable description of the $c$-axis charge
response in Sr$_{2}$RuO$_{4}$ and in particular the phonon dynamics
along this axis a typical LDA-based model is too isotropic and must be
modified to account for the much weaker interlayer coupling in the real
material. Similar as in the cuprates studied earlier the large
anisotropy in Sr$_{2}$RuO$_{4}$ is considerably underestimated in
DFT-LDA calculations of the electronic BS.

From our investigations we conclude that an accurate representation of
the very fainty $k_{z}$-dispersion of the BS nevertheless is essential
to understand the $c$-axis polarized $\Lambda_{1}$ modes. While the
LDA-based model (27BM) is insufficient a modified much more anisotropic
model (M27BM2) is well suited.

We have examined the multi-sheet FS of Sr$_{2}$RuO$_{4}$ with the two
models and found that the experimental results for the FS are not well
described within the 27BM. On the other hand, the M27BM2 shows a good
agreement with the measured FS. This model can be considered to
represent the real anisotropy in Sr$_{2}$RuO$_{4}$ sufficiently well.

We have also calculated the magnitude of the strongly enhanced
anisotropy in the M27BM2 compared with the 27BM more globally in terms
of some Fermi surface parameters important for transport properties.
Checking both models against each other we find an enhancement of about
a factor 18 for the anisotropy ratio of the Drude plasma energy tensor
and of about a factor of 10 for the Fermi velocity tensor. While the
calculations demonstrate that Sr$_{2}$RuO$_{4}$ is a nearly
two-dimensional Fermi liquid significantly more anisotropic than
La$_{2}$CuO$_{4}$ the remaining very weak three-dimensionality is
crucial to achieve a solid representation of the $c$-axis charge
response and of certain $c$-axis phonons.

We have calculated in detail the phonon dispersion in Sr$_{2}$RuO$_{4}$
along the main symmetry directions in the BZ and compared the result
with the dispersion of the HTSC La$_{2}$CuO$_{4}$ being structural
isomorphic. A good overall agreement of the calculated results is found
for Sr$_{2}$RuO$_{4}$ with the INS data. Comparing the results of the
calculated phonon dynamics in the RIM with a model allowing additional
DF's and finally the full model including DF's and CF's the
renormalization of certain modes which are strongly coupled via
nonlocal EPI effects of DF- and CF-type has been analyzed in detail. In
this context an important point of the comparison with
La$_{2}$CuO$_{4}$ is that the anomalous softening of the high-frequency
oxygen bond-stretching modes, being generic for the cuprate based
HTSC's, is strongly reduced or completely absent in Sr$_{2}$RuO$_{4}$
depending primarily on the magnitude of the on-site Coulomb repulsion
of the Ru4d orbitals.

We have also investigated the possibility of a Kohn anomaly in the
$\Sigma_{1}$ modes of Sr$_{2}$RuO$_{4}$ as discussed in the literature.
In our calculations we do not find any evidence for such an anomaly
driven by nesting of the FS. Instead, the dip in the lowest
$\Sigma_{1}$ branch seen in the experiments is well explained in our
computations by an anticrossing effect of the three lowest $\Sigma_{1}$
branches.

Finally, we have examined the question of a possible phonon-plasmon
scenario in a small region of nonadiabatic charge response around the
$c$-axis which has been shown to be a realistic option in
La$_{2}$CuO$_{4}$.

Due to the much weaker electronic $k_{z}$-dispersion obtained for
Sr$_{2}$RuO$_{4}$ the calculated free-plasmon frequencies along the
$c$-axis are about a factor of eight smaller in the collisionless
regime than in La$_{2}$CuO$_{4}$. We have argued that damping generated
by the interactions between the QP's as well as interband transitions
which are at much lower energy scale than in La$_{2}$CuO$_{4}$ leads to
an overdamping of the plasmon strictly along the $c$-axis in contrast
to La$_{2}$CuO$_{4}$. However, a coupled phonon-plasmon scenario
becomes likely also in Sr$_{2}$RuO$_{4}$ at higher free-plasmon
frequencies. This occurs in case the wavevector is not strictly
parallel to the $c$-axis but has a small transverse component. We find
that the strength of this nonlocal nonadiabatic coupling is
significantly weaker than in La$_{2}$CuO$_{4}$.

Ultimately we have explained by our calculations of the nonadiabatic
charge response the linewidth of the apex oxygen breathing mode at the
$Z$ point and why $c$-axis optical activity as seen in the experiments
is possible despite the fact that Sr$_{2}$RuO$_{4}$ is in the metallic
phase.



\section*{References}
\begin{thebibliography}{10}
\bibitem{Maeno94} Maeno Y et al. 1994 Nature \textbf{372} 532
\bibitem{Bergemann03} Bergemann C, MacKenzie AP, Julian SR, Forsythe D and Ohmichi E 2003 Adv. Phys. \textbf{52} 639
\bibitem{Schofield05} Ho AF and Schofield AJ 2005 Phys. Rev. B \textbf{71} 045101
\bibitem{Mackenzie03} Mackenzie AP, Maeno Y 2003 Rev. Mod. Phys. \textbf{75} 657
\bibitem{Mao03} Mao ZQ et al. 2003 Phys. Rev. B \textbf{63} 144514
\bibitem{Falter05} Falter C 2005 Phys. Status Solidi B \textbf{242} 78
\bibitem{Bauer08} Bauer T and Falter C 2008 Phys. Rev. B \textbf{77} 144503
\bibitem{Falter06} Falter C, Bauer T, and Schnetg\"oke F 2006 Phys. Rev. B \textbf{73} 224502
\bibitem{Pint05} Pintschovius L 2005 Phys. Status Solidi B \textbf{242} 30
\bibitem{Falter93} Falter C, Klenner M and Ludwig W 1993 Phys. Rev. B \textbf{47} 5390
\bibitem{Braden07} Braden M, Reichardt W, Sidis Y and Maeno Y 2007 Phys. Rev. B \textbf{76} 014505
\bibitem{Bauer09} Bauer T and Falter C 2009 cond-mat/0808.2765
\bibitem{Falter99} Falter C, Klenner M, Hoffmann GA and Schnetg\"oke F 1999 Phys. Rev. B \textbf{60} 12051
\bibitem{Falter88} Falter C 1988 Phys. Rep. \textbf{164} 1
\bibitem{Falter95} Falter C, Klenner M and Hoffmann GA 1995 Phys. Rev. B \textbf{52} 3702
\bibitem{Falter02} Falter C and Schnetg\"oke F 2002 Phys. Rev. B \textbf{65} 054510
\bibitem{Perdew81} Perdew JP and Zunger A 1981 Phys. Rev. B \textbf{23} 5048
\bibitem{Krakauer98} Krakauer H, Pickett WE and Cohen RE 1998 J. Supercond. \textbf{1} 11
\bibitem{Savrasov96} Savrasov SY and Andersen OK 1996 Phys. Rev. Lett. \textbf{77} 4430
\bibitem{Wang99} Wang CZ, Yu R and Krakauer H 1999 Phys. Rev. B \textbf{59} 9278
\bibitem{Bohnen03} Bohnen KP, Heid R and Krauss M 2003 Europhys. Lett. \textbf{64} 104
\bibitem{Giustino08} Giustino F, Cohen ML and Louie SG 2008 Nature \textbf{452} 975
\bibitem{Mazin00} Mazin II, Papaconstantopoulos DA and Singh DJ 2000 Phys. Rev. B \textbf{61} 5223
\bibitem{Shen07} Shen KM et al. 2007 Phys. Rev. Lett. \textbf{99} 187001
\bibitem{Chmaissem98} Chmaissem O et al. 1998 Phys. Rev. B \textbf{57} 5007
\bibitem{Moore08} Moore RG et al. 2008 Phys. Rev. Lett. \textbf{100} 066102
\bibitem{Uruma07} Uruma M et al. 2008 cond-mat/0711.2160 (unpublished)
\bibitem{Graf08} Graf J, et al. 2008 Phys. Rev. Lett. \textbf{100} 227002
\bibitem{Sidis99} Sidis J et al. 1999 Phys. Rev. Lett. \textbf{83} 3320
\bibitem{Braden02} Braden M et al. 2002 Phys. Rev. B \textbf{66} 064522
\bibitem{Tamasaku92} Tamasaku K, Nakamura Y and Uchida S 1992 Phys. Rev. Lett. \textbf{92} 1455
\bibitem{Pintpc} Pintschovius L, private communication
\bibitem{Katsufuji96} Katsufuji T, Kasai M and Tokura Y 1996 Phys. Rev. Lett. \textbf{76} 126
\end{thebibliography}
\end{document}